\documentclass{article}
\usepackage[a4paper, margin=1in]{geometry}
\usepackage{authblk}
\usepackage{fontenc}[T1]
\usepackage{alltt}
\usepackage{graphicx}
\usepackage[toc, page]{appendix}

\usepackage[english]{babel}
\usepackage{multirow}
\graphicspath{{Figures/}}
\usepackage{hyperref}
\usepackage{amsthm}
	
\usepackage{enumitem}
\usepackage{multirow}
\usepackage{booktabs}
\usepackage{amsmath}
\usepackage{amsfonts}
\usepackage{amssymb}
\usepackage{hyperref}
\usepackage{mdwlist}
\usepackage[lined, boxruled]{algorithm2e}
\usepackage{color}
\usepackage{pgf}
\usepackage{longtable}
\usepackage{pdflscape}
\usepackage[round]{natbib} 
\usepackage{mathrsfs}
\usepackage[font={small}]{caption}
\usepackage{subcaption}
\usepackage{graphicx}
\usepackage{rotating}
\usepackage{mathtools}
\usepackage[textsize=footnotesize]{todonotes}
\usepackage{bbm}

\newtheorem{theorem}{Theorem}
\newtheorem*{theorem*}{Theorem}
\newtheorem{lemma}[theorem]{Lemma}

\newtheorem{definition}[theorem]{Definition}

\newcommand{\cyclen}{K}
\newcommand{\chainlen}{L}
\newcommand{\C}{\mathcal{C}}
\newcommand{\D}{\mathcal{D}}
\newcommand{\E}{\mathcal{E}}
\renewcommand{\P}{\mathcal{P}}
\newcommand{\X}{\mathcal{X}}
\newcommand{\N}{\mathbb{N}}

\newcommand{\true}{\texttt{True}}
\newcommand{\false}{\texttt{False}}

\begin{document}

\title{Rejection-proof Kidney Exchange Mechanisms}
\author{Blom D.\thanks{Corresponding author}, Smeulders B., Spieksma F.C.R.}
\affil{Eindhoven University of Technology}
\date{\today}

\maketitle

\begin{abstract}
    Kidney exchange programs (KEPs) form an innovative approach to increasing the donor pool through allowing the participation of renal patients together with a willing but incompatible donor. The aim of a KEP is to identify groups of incompatible donor-recipient pairs that could exchange donors leading to feasible transplants. As the size of a kidney exchange grows, a larger proportion of participants can be transplanted. Collaboration between multiple transplant centers, by merging their separate kidney exchange pools is thus desirable. As each transplant center has its own interest to provide the best care to its own patients, collaboration requires balancing individual and common objectives. We consider a class of algorithmic mechanisms for multi-center kidney exchange programs we call \emph{rejection-proof mechanisms}. Such mechanisms propose solutions with the property that no player wishes to unilaterally deviate. We provide a mechanism optimizing social value under this restriction, though the underlying optimization problem is $\Sigma_2^p$-hard. We also describe a computationally easier but sub-optimal alternative. Experiments show that rejection-proofness can be achieved at limited cost compared to optimal solutions for regular kidney exchange. Computationally, we provide algorithms to compute optimal rejection-proof solutions for small and medium instance sizes.

\end{abstract}

\section{Introduction}
\begin{figure}[ht!]
    \centering
    \begin{tikzpicture}
\node [rectangle, draw] (v1) at (0, 0) {1};
\node [circle, draw] (v2) at (2, 0) {2};
\node [circle, draw] (v3) at (4, 0) {3};
\node [circle, draw] (v4) at (0, -2) {4};
\node [circle, draw] (v5) at (2, -2) {5};
\node [circle, draw] (v6) at (4, -2) {6};

\draw [-latex] (v1) edge[bend left = 00, ultra thick] (v2);
\draw [-latex] (v2) edge[bend left = 00] (v3);
\draw [-latex] (v2) edge[bend left = 00, ultra thick] (v4);
\draw [-latex] (v2) edge[bend left = 00] (v6);
\draw [-latex] (v3) edge[bend left = 10, ultra thick] (v5);
\draw [-latex] (v5) edge[bend left = 10, ultra thick] (v3);
\draw [-latex] (v4) edge[bend left = 10] (v5);
\draw [-latex] (v5) edge[bend left = 10] (v4);
\draw [-latex] (v6) edge[bend left = 10] (v5);
\draw [-latex] (v5) edge[bend left = 10] (v6);

\end{tikzpicture}
    \caption{Circles represents recipient-donor {\em pairs}, while squares represent {\em non-directed donors}. The presence of an arc $(i,j)$ signifies compatibility of the donor of $i$ and the recipient of $j$. The bold arcs represent a feasible solution, with a cycle between pairs 3 and 5 and a chain from NDD 1 to pair 2 and on to pair 4.}
    \label{fig:cyclechain}
\end{figure}
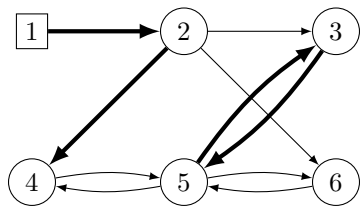

For patients suffering from end-stage renal disease, a kidney transplant from a living donor is the preferred treatment option. A donor must be medically compatible with the recipient, which means some recipients can not receive a transplant, even though they have a donor willing to undergo a transplant for them. We refer to such a combination of recipient and donor as a {\em pair}. Kidney exchange programs (KEPs), introduced by \cite{Rapaport1986} and popularized by \cite{Roth2004,Roth2005} aim to match the recipients from a pair with compatible donors from other pairs. Donors in a pair only donate if their paired recipient receives a transplant in return, leading to cycles of transplants. Kidney exchange programs may also include so-called non-directed donors (NDDs). These donors are not associated with a recipient, and are willing to donate to any recipient in the program, leading to donation chains. Typically, kidney exchanges have a policy limiting the maximum length of cycles and chains. Figure \ref{fig:cyclechain} illustrates a kidney exchange graph and solution. Kidney exchange programs are widely used, and are still expanding. We refer to \cite{biro2019} for a recent overview of the practice in Europe.

\begin{figure}[ht]
    \centering
    \includegraphics[scale=0.8]{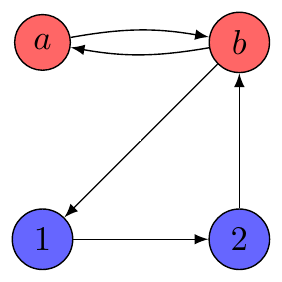}
    \caption{While the social optimum of this 2-agent KEP instance is given by $c = (b, 1, 2)$, the red agent can choose to withholding both its pairs, thereby gaining one more transplant for its own recipients.}
    \label{fig:cooploss}
\end{figure}

Kidney exchanges with larger pools of pairs and NDDs generally provide better outcomes for all participants, as the chances of identifying compatible donors increase. \cite{agarwal2020market} show that many hospital and regional kidney exchange programs in the US operate well below an efficient size. Within Europe, several countries have started transnational cooperations (known as \emph{multi-agent kidney exchange programs}) to increase pool size \citep{bohmig2017czech, valentin2019international}. While merging separate kidney exchange programs is an appealing idea, cooperation also faces hurdles. These vary from legal considerations and the alignment of medical practices, but also from the risk of unequal sharing of benefits. In extreme cases, unrestricted cooperation may even lead to a loss of transplants for an individual hospital or country involved in the merged pool, as illustrated in Figure \ref{fig:cooploss} (which is taken from \cite{ashlagi2014free}. We will use the term {\em agent} to refer to an entity (hospital or country) that controls a set of pairs and non-directed donors, and we will use the term {\em collaboration} to refer to a central authority that is able to propose solutions, i.e., transplant plans that contain pairs or non-directed donors from multiple agents.

The example in Figure \ref{fig:cooploss} shows that maximizing transplants in merged pools gives incentives for strategic behaviour. Within the literature on kidney exchange, this behaviour concerns the sharing and withholding of information. Typically, agents choose which of their pairs and NDD's to reveal to the collaboration. The collaboration then proposes a transplant plan, matching some of the agent's recipients. The agent can then also select transplants for their unmatched (either hidden, or revealed but not part of the collaboration's transplant plan) pairs \citep{ashlagi2014free}. We will refer to this behaviour by the agents by \textit{withholding strategies}.
In particular, it can be shown that no socially optimal mechanisms that are strategy-proof exist \citep{sonmez2013market, ashlagi2014free}. Work on cooperation in kidney exchange has focused on mechanisms to incentivize agents to reveal their complete pool, mostly through credit mechanisms in multi-period settings (\cite{hajaj2015strategy},  \cite{biro2019}, \cite{agarwal2020market}, \cite{biro2020}, \cite{biro2020compensation}, \cite{klimentova2020fairness}) or randomized mechanisms (\cite{ashlagi2015mix}).  \\

\begin{figure}[ht]
    \centering
    \includegraphics[scale = 0.7]{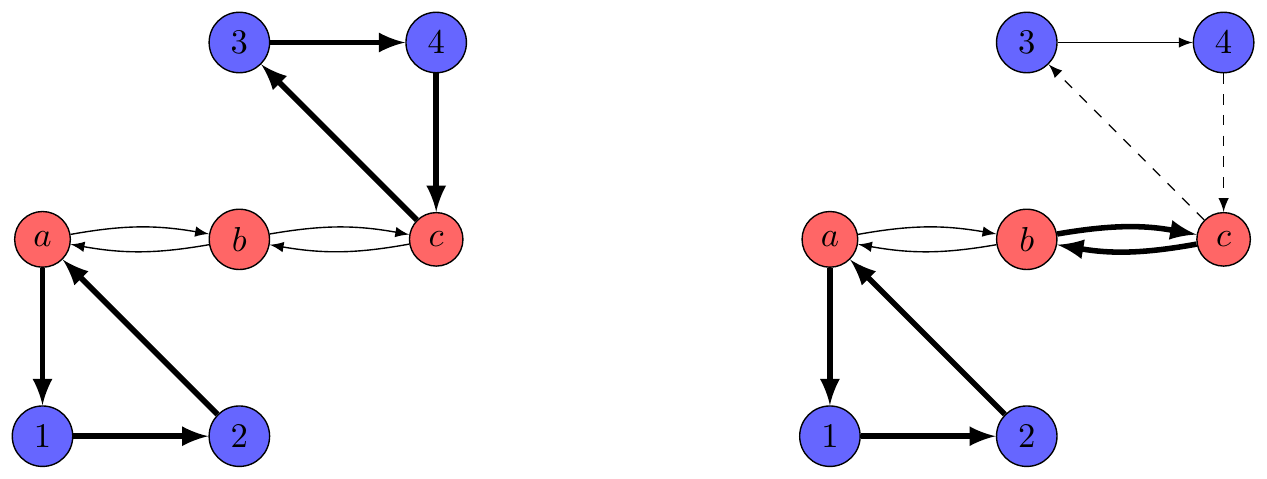}
    \caption{The social optimum (left) consists of two 3-cycles $(a,1,2)$ and $(c,3,4)$. If this is the proposed solution, the red agent has an incentive to reject the cycle $(c,3,4)$ and internally match $(b,c)$, thereby increasing the number of transplants for the red agent.}
    \label{fig:partialreject}
\end{figure}

In this paper, we consider a different setting, where play use what we will call \textit{rejection strategies}. Here, we assume the collaboration has full information about agent pools, and the collaboration proposes a solution. Next, agents decide whether to reject (part of) this proposed solution. This entails withdrawing pairs or NDDs from cycles and chains from the proposed solution. Afterwards, the agents can internally select new cycles and chains with pairs or NDDs that were either withdrawn or unmatched in the proposed solution. An example is given in Figure \ref{fig:partialreject}. \\

Rejection strategies are relevant when the collaboration cannot enforce the proposed solution. This is actually the case in a kidney exchange setting, as binding agreements to execute a proposed transplant plan are impossible. Indeed, any measure forcing patients and donors to undergo surgery are legally and ethically unacceptable. Consequently, any solution proposed by the collaboration may be impacted by (unforeseen) withdrawals. Moreover, from the point of view of the collaboration, it may be unclear what the precise cause of a withdrawal of a pair of a particular agent is – it is conceivable that the withdrawn pair is used in a transplant that is hidden from the sight of the collaboration perhaps in order to increase the number of transplants for an individual agent. All this motivates a concept that we study in this contribution: the concept of solutions that are rejection-proof (see Section 2 for formal definitions). The goal of this paper is to study mechanisms that lead to solutions such that no agent has an incentive to withdraw a pair or NDD from the proposed solution.

The paper unfolds as follows. In Section \ref{sec:Prelim}, we introduce necessary notation, we formally define withholding and rejection strategies, and we introduce the {\em rejection-proof} property. In Section \ref{sec:WithvsReject} we experimentally compare withholding strategies and rejection strategies by considering the resulting loss in total transplants for these strategies. Sections \ref{sec:RPKEP} and \ref{sec:MaxIntMech} introduce two rejection-proof mechanisms. The first mechanism, Maximum Rejection-proof KEP (MaxRP-KEP), always proposes the maximum value rejection-proof solution. However, we show finding this solution is a $\Sigma_2^p$-hard problem and the mechanism thus requires solving a complex bilevel optimization problem. The second mechanism, Maximum Internal Transplants (MaxInt) is inspired by current practice. The mechanism is intuitive, and relatively simple from a computational point of view, but proposes sub-optimal solutions. In Section \ref{sec:CompResults}, we describe the outcomes of the two mechanisms. We show that MaxRP-KEP achieves rejection-proofness at a negligible cost in terms of transplants. Furthermore, it turns out that both mechanisms reduce the benefits of withholding strategies as well. We round off the paper with a discussion in Section \ref{Sec:Disc}.

\section{Preliminaries and Notation} \label{sec:Prelim}
Consider a directed graph $G = (V,A)$ and a set of agents $\P$. The vertex set is partitioned into subsets $V^i$ for each agent $i \in \P$, i.e., $V = \dot\bigcup_{i \in \P} V^i$. Furthermore, for each $i \in \P$, the set $V^i$ is divided into a set of non-directed donors $N^i$ and a set of pairs $P^i$, i.e. $V^i = N^i \dot\cup P^i$. Let $\C$ be the set of cycles which can be used for transplants (typically all cycles of length at most $K$), and for each $i \in \P$, let $\C^i \subseteq \C$ be the subset of these cycles on the induced subgraph $G^i = G[V^i]$. We refer to $\C^i$ as the set of {\em internal} cycles of agent $i$, $(i \in \P)$. Furthermore, we have $\C^0 = \C \setminus \bigcup_{i \in \P} \C^i$, which consists of cycles containing pairs of multiple agents, referred to as {\em shared} cycles. Let $w_c$ be the value of the cycle, and $w^i_c$ the value of the cycle to agent $i \in \P$. For chains, we restrict ourselves to transplants between donors and recipients belonging to the same agent. The donor of the final pair in a chain can, and often does, become a NDD in later matching runs, which leads to additional transplants in the long run. For this reason, we assume agents prefer to keep chains internal, so they have more NDDs in the following matching runs. Let $\D^i$ be the set of all allowed chains of an agent $i \in \P$ (typically all acyclic paths of length at most $L$ in $G^i$ originating in a node $v \in N^i$). We denote by $\D = \bigcup_{i \in \P} \D^i$ the set of allowed chains in $G$. Again, $w_d$ denotes the value of the chain $d$, and $w^i_d$ the value to agent $i$. For the sake of brevity, we will use the term {\em exchange} to refer to either a cycle or a chain, and define the set of exchanges $\E$ by setting $\E = \C \cup \D$. For each agent $i \in \P$, we use $\E^i = \C^i \cup \D^i$ to denote the set of internal exchanges to agent $i$ and shared exchanges are denoted by $\E^0$. For each exchange $e \in \E$, we denote by $V(e)$ and $A(e)$ its sets of vertices and arcs respectively.\\

Typically, the values $w_c$ ($w_c^i$) and $w_d$ ($w_d^i$) are considered to be the number of transplants in the cycle or chain (to recipients controlled by agent $i$). However, note that we allow $\sum_{i \in \P} w^i_c \ne w_c$, for example, the participating agents may have agreed that the cooperation must strive for the maximum number of transplants, but privately wish to prioritize certain transplants.

The set of feasible kidney exchange solutions is captured by the following set
\[
    \X = \{X \subseteq \E : V(e)\cap V(e') = \emptyset\quad \forall e, e' \in X, e \ne e' \ \}.
\]
Note that for each $X \in \X$, the set of cycles and chains forming a solution, the vertex sets of the exchanges must be disjoint. This condition is required, as donors can donate at most one kidney and each pair or NDD can thus be involved in at most one exchange. We denote the value of a solution $X \in \X$ by $w(X) = \sum_{e \in X} w_e$, the weight of all selected cycles and chains. The value of a solution to a particular agent is denoted by $w^i(X) = \sum_{e \in X} w^i_e$. Furthermore, let us denote by $V(X)$ the set of all pairs and NDDs matched with respect to $X$.\\

The challenge in multi-agent kidney exchange problems is to settle upon a satisfying way to agree upon the interplay between the agents on the one hand, and the collaboration on the other hand. In other words, how to design a mechanism to arrive at a solution; we define a mechanism as follows:
\begin{definition}
A kidney exchange mechanism is a function $f: G \rightarrow \X$ which, given a graph $G = (V,A)$ and its partition of $V$ into subsets $V^i$ for each $i \in \P$, produces a kidney exchange solution $X = f(G) \in \X$.
\end{definition}

\subsection{Withholding strategies} \label{sec:withholding}
The traditional setting, in which only withholding strategies are used, implies the following game:
\begin{enumerate}
    \item Each agent $i \in \P$ identifies a subset $S^i \subseteq V^i$ of its pairs and NDDs to contribute to the shared kidney exchange pool. Throughout the rest of the paper, we will refer to this as agent $i \in \P$ selecting the withholding strategy $W^i = V^i \setminus S^i$.
    \item A mechanism $f$ computes a solution $X = f(G[\bigcup_{i \in \P} S^i])$. We denote by $U^i(X) \subseteq S^i$ the set of contributed pairs and NDDs of agent $i \in \P$ that remain unmatched with respect to $X$.
    \item Each agent $i \in \P$ selects a maximum weight kidney exchange solution $X^i$ on the subgraph of $G^i$ induced by $W^i \cup U^i(X)$.
\end{enumerate}

For each agent $i \in \P$, the value is given by $w^i(X \cup X^i)$. We note that it is computationally hard for an agent to maximize its value. Indeed,~\cite{Smeulders2022Stackel} show that the problem of finding an optimal withholding strategy for agent $i \in \P$ is $\Sigma_2^p$-hard, even when the mechanism $f$ outputs a social optimal kidney exchange solution and the withholding strategies $W^j \subseteq V^j$ of each agent $j \in \P \setminus \{i\}$ are fixed and known to agent $i$. This difficulty is amplified by the fact that in practice the vertex sets $V^j$ of other agents are not known to agent $i$, so they are not even aware of the possible withholding strategies other agents could select. \\

Due to the complexity of computing optimal withholding strategies, we consider an intuitive withholding strategy proposed in \cite{ashlagi2014free}, which we call the \emph{greedy withholding strategy}. With this strategy, agent $i \in P$ selects $W^i \subseteq V^i$, with $W^i$ the set of pairs and NDDs matched in a kidney exchange solution of maximum weight to agent $i$ using only the exchanges internal to agent $i$. The vertices $S^i = V^i \setminus W^i$ are then contributed to the shared pool. In a situation where all other agents submit all their pairs and NDDs to the collaboration, \cite{ashlagi2014free} show experimentally that the greedy withholding strategy increases, on average, the number of transplants for an individual agent employing this strategy.

\subsection{Rejection strategies}
In this paper, we consider a setting that requires a mechanism to propose a solution first, after which the individual agents can select a \emph{rejection strategy}. This setting assumes that the entire compatibility graph $G$ is known to the mechanism, and not necessarily to the individual agents.
\begin{enumerate}
    \item A mechanism $f$ selects a solution $X = f(G)$.
    \item Each agent $i \in \P$ selects a rejection strategy, i.e. a kidney exchange solution $X^i \in \X$ such that for all exchanges $e \in X^i$, we have that $V(e) \cap V^i \ne \emptyset$ and $(X^i \cap \E^0) \subseteq (X \cap \E^0)$.
    \item A solution $X' \in \X$ is found where $e \in X'$ if and only if $e \in X^i$ for every $i \in \P$ for which $V(e) \cap V^i \ne \emptyset$.
\end{enumerate}
We say that agent $i \in \P$ \emph{accepts} a proposed solution $X$ if agent $i$ selects the rejection strategy $X^i = \{e \in X \mid V(e) \cap V^i \ne \emptyset\}$, i.e., $X^i$ consists of all exchanges in the proposed solution $X$ involving at least one pair or NDD of $V^i$. If this is not the case, then we say that agent $i$ \emph{rejects} the solution.

The goal of an agent $i \in \P$ is to select a rejection strategy $X^i$ that has maximum value to agent $i$ itself. We call this problem the Reject Kidney Exchange Problem (RKEP). A mixed-integer programming formulation RKEP$(X, i)$ of this optimization problem is described in~\eqref{IP:rejectproblem}.
It considers a binary variable $x^i_e$ for each exchange $e \in \E$, which has value 1 if and only if $e$ is selected in the rejection strategy $X^i$. The agent is free to choose any set of pairwise disjoint internal exchanges ($\E^i$), while Constraints~\eqref{IP:reject:shared} imply that shared exchanges $e \in \E^0$ can only be included if the exchange is contained in the proposed solution $X$.

\begin{subequations} \label{IP:rejectproblem}
\begin{align}
    \text{RKEP}(X, i) := \text{maximize} \quad \sum_{e \in \E} w^i_e x^i_e \ && \label{Model: Cycle Country Objective}\\
     \text{subject to} \quad \sum_{e \in \E: v \in V(e)} x^i_e \ & \le \ 1 & \forall v \in V \\
   x^i_e \ & = \ 0 & \forall e \in \E^0 \setminus X \label{IP:reject:shared} \\
   x^i_e \ &  \in \ \{0,1\} & \forall e \in \E \label{Model: Cycle Country Binary}
\end{align}
\end{subequations}

Notice that for any agent $i \in \P$ for which RKEP$(X, i) \le w^i(X)$, it is a weakly dominant strategy to accept the proposed solution, provided that the other agents also accept the solution. If all agents accept the solution $X$, each agent $i \in \P$ scores a value of $w^i(X)$. On the other hand, if for some agent $i \in \P$, RKEP$(X, i) > w^i(X)$, agent $i$ can achieve a superior outcome by rejecting the proposed solution. An optimal rejection strategy is then encoded by the $x^i_e$ variables. Remark however that a rejection strategy $X^i$ does not guarantee that agent $i$ receives a payout of $w^i(^i)$, as there might exist a shared exchange $e \in \E^0 \cap X$ not contained in the rejection strategy of another agent $j$ having a vertex involved in $e$.

\subsection{Rejection-proofness}
In this work, we will consider mechanisms that propose solutions $X$, for which accepting is a weakly dominant strategy for every agent, assuming other agents also accept. This gives rise to the definition of \textit{rejection-proofness}.

\begin{definition}[Rejection-proof]
A mechanism $f: G \rightarrow \X$ is rejection-proof if, for any compatibility graph $G = (V,A)$ and each agent $i \in \P$, we have
\begin{align}
    \emph{RKEP}(f(G), i) \le w^i(f(G))
\end{align}
\end{definition}

Informally, we will also say that an individual solution $X$ is rejection-proof if and only if RKEP$(X, i) \le w^i(X)$ for all agents $i \in \P$. Note that in effect, any proposed solution $X$ gives rise to a game between the agents. $X$ determines the set of rejection strategies for each agent, and $w^i(X')$ defines the value to agent $i$. Observe that by definition, for a rejection-proof mechanism, it holds that the profile of rejection strategies $(X^1, \ldots, X^{|\P|})$ given by $X^i = X \cap \{e \mid V(e) \cap V^i \ne \emptyset\}$, i.e., each agent accepts the solution $X$, is a pure Nash equilibrium of this game.

\section{Experimentally comparing withholding and rejection strategies}\label{sec:WithvsReject}

In the previous section, we have formally defined both withholding and rejection strategies, as well as the rejection-proofness property. In this section, we experimentally show the benefits agents can achieve through using these strategies, as well as the loss in social value incurred by this strategic behaviour. This will show that incentives for strategic behaviour are present in realistic settings, and that the resulting overall loss is large enough to motivate the development of mechanisms preventing strategic behaviour.

We use the same experimental set-up as \cite{ashlagi2014free}. These authors evaluated the gains agents could achieve through the greedy withholding mechanism. In this section, we extend those simulations to also evaluate rejection strategies.

We consider KEP instances generated through two sets of characteristic distributions for donor-recipient pairs. One set is generated using the traditional \cite{saidman2006increasing} distributions (also used in Ashlagi and Roth). The second set is constructed based on recent work, which aims to more closely resemble instances of the UK's KEP, as described by \cite{delorme2021new} Specifically, we make use of the SplitPRA-BandXMatch-PRA0 distributions described in this paper. Both sets of instances were generated using the generator found at \url{https://wpettersson.github.io/kidney-webapp/#/generator}, making use of the pre-built configurations. We consider the same instance sizes as Ashlagi and Roth. Specifically, we create instances with an equal number of pairs for all agents, either 10, 15, 30, 50 or 80. The number of agents is always even and chosen such that there are at most 180 total pairs. We assume agents value every transplant equally, and that there are no non-directed donors.

Results of the experiments are described in Table \ref{tab:agentStrat}. This table shows the average gain for an individual agent employing the greedy withholding or optimal rejection strategy, assuming all other agents do not employ such strategies. We note that for both instance types, withholding and rejection strategies provide similar benefits for agents with small pools. For larger agents, with pool sizes of 30 and above, the average gain for the rejection strategy is consistently larger than for the withholding strategy. Furthermore, we note that even though in all but one case, the average gain of withholding was positive, nearly every instance set contains instances where withholding led to transplant losses (only DelBXMPRA0\_10x8 and DelBXMPRA0\_10x14 did not). By contrast, rejection strategies are risk-free to the agent. If an agent fails to identify a rejection strategy with which the payout can be improved over accepting, the agent can keep the proposed solution. The dual result of higher average profits and reduced risk makes rejection strategies a superior choice for agents.

\begin{table}[htbp]
    \centering\footnotesize
    \begin{tabular}{cccccccccc}
        \toprule
        Instance set & \multicolumn{4}{c}{Delorme SplitPRA-BXM-PRA0} & & \multicolumn{4}{c}{Saidman}\\
        \midrule
        & WA &  RA & WT & RT & & WA & RA & WT & RT\\
        \midrule
        10x10 &          1.005 &         1.005 &                     0.943 &                    0.986 &&          1.003 &         1.034 &                     0.868 &                    0.931\\
        10x12 &          1.012 &         1.008 &                     0.931 &                    0.985 &&         1.040 &         1.037 &                     0.890 &                    0.929 \\
        10x14 &          1.019 &         1.012 &                     0.944 &                    0.985 &&          1.021 &         1.050 &                     0.900 &                    0.925 \\
        10x16 &          1.019 &         1.004 &                     0.938 &                    0.988 &&          1.031 &         1.046 &                     0.899 &                    0.921 \\
        10x2  &          1.000 &         1.000 &                     0.935 &                    1.000 &&          0.992 &         1.042 &                     0.855 &                    0.981 \\
        10x4  &          1.045 &         1.055 &                     0.882 &                    0.962 &&          1.038 &         1.064 &                     0.846 &                    0.943 \\
        10x6  &          1.014 &         1.014 &                     0.926 &                    0.986 &&          1.068 &         1.065 &                     0.844 &                    0.929 \\
        10x8  &          1.006 &         1.006 &                     0.927 &                    0.980 &&          1.066 &         1.066 &                     0.869 &                    0.907 \\
        15x10 &          1.015 &         1.010 &                     0.905 &                    0.977 &&          1.051 &         1.060 &                     0.886 &                    0.914 \\
        15x2  &          1.000 &         1.008 &                     0.939 &                    0.991 &&          1.028 &         1.077 &                     0.862 &                    0.961 \\
        15x4  &          1.009 &         1.017 &                     0.873 &                    0.971 &&          1.076 &         1.100 &                     0.846 &                    0.915 \\
        15x6  &          1.046 &         1.025 &                     0.877 &                    0.969 &&          1.047 &         1.087 &                     0.860 &                    0.915 \\
        15x8  &          1.024 &         1.038 &                     0.907 &                    0.975 &&          1.020 &         1.047 &                     0.867 &                    0.909 \\
        30x2  &          1.004 &         1.055 &                     0.854 &                    0.981 &&          1.040 &         1.080 &                     0.901 &                    0.959 \\
        30x4  &          0.992 &         1.039 &                     0.813 &                    0.955 &&          1.044 &         1.077 &                     0.871 &                    0.912 \\
        30x6  &          1.012 &         1.036 &                     0.839 &                    0.958 &&          1.052 &         1.088 &                     0.881 &                    0.894 \\
        50x2  &          1.003 &         1.074 &                     0.820 &                    0.966 &&          1.032 &         1.055 &                     0.928 &                    0.950 \\
        80x2  &          0.984 &         1.059 &                     0.807 &                    0.952 &&          1.033 &         1.058 &                     0.958 &                    0.955 \\
        \bottomrule
    \end{tabular}\footnotesize
    \caption{ Comparison of withholding and rejection strategies. {\bfseries Instance Set:} The first number of the instance set shows the number of pairs each agent has, the second shows the number of players. {\bfseries Withhold Agent (WA):} average gain or loss in number of transplants for a specific agent from switching unilaterally from no withholding to greedy withholding, assuming all other agents do not withhold. {\bfseries Reject Agent (RA):} similar, but with rejection strategies. {\bfseries Withhold Total (WT):} average gain or loss in the total number of transplants when comparing full cooperation of all agents to greedy withholding strategies for all agents. {\bfseries Reject Total (RT):} similar, but with rejection strategies for all agents.}
    \label{tab:agentStrat}
\end{table}

The table also shows the loss of transplants overall, if all agents employ the relevant strategy. In our experiments, rejection strategies can lower the total number of transplants by up to 10\%. While this is lower than the losses caused by the greedy withholding strategy, they are not negligible. As a result, we conclude that mechanisms should also address the risk of rejection strategies.

\section{Maximum rejection-proof kidney exchange solutions}\label{sec:RPKEP}
In this section, we will formulate a mechanism that proposes a rejection-proof solution of maximum weight, dubbed the \textit{Maximum Rejection-Proof KEP} mechanism or simply MaxRP-KEP. We also propose an algorithm for solving the resulting bilevel optimization problem.

\begin{definition}{Maximum Rejection-Proof KEP (MaxRP-KEP)}\\
The MaxRP-KEP mechanism selects a kidney exchange solution $X = (x_e)_{e \in \E}$ optimal to the following bilevel program.
\begin{subequations} \label{IP:MaxRP-KEP}
\begin{align}
    \emph{maximize} \ \sum_{e \in \E} w_e x_e \ && \label{IP:MaxRP-KEP:obj}\\
    \emph{subject to} \ \sum_{e \in \E: v \in V(e)} x_e \ & \le \ 1 && \forall v \in V \label{IP:MaxRP-KEP:packing}\\
    \sum_{e \in \E} w^i_e x_e \ & \ge \ \emph{RKEP}(X, i) && \forall i \in \P \label{IP:MaxRP-KEP:agentAccept}\\
    x_e \ & \in \ \{0,1\} && \forall e \in \E  \label{IP:MaxRP-KEP:binary}
\end{align}
\end{subequations}
\end{definition}

If we restrict ourselves to the objective~\eqref{IP:MaxRP-KEP:obj} and Constraints~\eqref{IP:MaxRP-KEP:packing}, we have a standard weighted kidney exchange problem. Constraints~\eqref{IP:MaxRP-KEP:agentAccept} enforce that no agent has a strict incentive to reject the solution. Adding these constraints seriously complicates the problem, as the right-hand is the value to an agent for their optimal rejection strategy. Indeed, Model~\eqref{IP:MaxRP-KEP} is actually a bilevel program. We show that for cases in which standard kidney exchange problems are NP-hard, the problem of finding maximum cardinality rejection-proof kidney exchange solutions is complete in the complexity class $\Sigma_2^p$ (see \cite{woeginger2021trouble} for more background).

\begin{theorem}\label{th:complexity_proof}
Given a set of agents $\P$, a graph $G = (V,A)$ with a partition of $V$ into subsets $V^i$ for each agent $i \in \P$, a fixed maximum cycle length $K \ge 3$, any maximum chain length $L \in \N$ and an integer $t$, deciding whether there exists a rejection-proof solution $X \in \X$ with value at least $t$ is $\Sigma_2^p$-complete.
\end{theorem}
The proof of this theorem can be found in the Appendix. Solving $\Sigma_2^p$-hard problems is challenging, and requires specialized algorithms. In the remainder of this section, we discuss strategies to solve Model~\eqref{IP:MaxRP-KEP}. In Section~\ref{sec:reject constraints}, we propose a single-level reformulation of the problem with exponentially many constraints. Next, Section~\ref{sec:Solving} discusses an iterative constraint generation procedure.

\subsection{A single-level reformulation}\label{sec:reject constraints}

In this section, we identify a set of constraints which can replace~\eqref{IP:MaxRP-KEP:agentAccept} in Model~\eqref{IP:MaxRP-KEP}, thus reducing it to a single-level optimization problem, albeit of size exponential in the size of the input. First, for each agent $i \in \P$ and each subset $U^i \subseteq V^i$, we define
\begin{align}
    \beta(U^i) = \max\left\{\sum_{e \in X} w^i_e \mid X \in \X: \forall e \in X, V(e) \subseteq U^i\right\}.
\end{align}
$\beta(U^i)$ denotes the maximum value to agent $i$ of feasible kidney exchange solutions on $G[U^i]$. Note that calculating $\beta(U^i)$ requires solving a standard kidney exchange problem, a NP-hard problem. We now claim the following:

\begin{theorem} \label{Theorem: Stability cycle Form}
A solution $X = (x_e)_{e \in \E}$ to \emph{MaxRP-KEP} is rejection-proof if and only if for each agent $i \in \P$ and each set $U^i \subseteq V^i$, we have
\begin{align}
    \sum_{e \in \E: V(e) \cap U^i \ne \emptyset} w^i_e x_e  \ge \beta(U^i). \label{IS Conditions}
\end{align}
\end{theorem}
\begin{proof}
$\Rightarrow$ Suppose there exists some agent $i \in \P$ and a subset $U^i \subseteq V^i$ for which the solution $X$ does not satisfy~\eqref{IS Conditions}. Then agent $i$ can improve their solution by rejecting all exchanges containing a node $v \in U^i$ and replacing them by a kidney exchange solution on $G[U^i]$ with maximum value to the agent. Hence $X$ is not rejection-proof.\\

$\Leftarrow$ Suppose the solution $X = (x_e)_{e \in \E}$ is not rejection-proof. Then there exists an agent $i \in \P$ and a solution $X^i = (x^i_e)_{e \in \E} \subseteq X \cup \E^i$ feasible to $\eqref{IP:rejectproblem}$ such that
\begin{align}
    \sum_{e \in \E} w^i_e x^i_e  > \sum_{e \in \E} w^i_e x_e. \label{IS Conditions Proof 1}
\end{align}
We now construct the set $U^i$ as follows:
\begin{align}
    U^i := \{v \in V^i \mid \exists e \in \E: v \in V(e), x_e^i = 1, x_e = 0 \}.
\end{align}
Thus, the set $U_i$ is the set of vertices that are part of an exchange in $X_i$, and not part of an exchange in $X$. By construction of the set $U^i$, we have that $x^i_e \le x_e$ if $V(e) \cap U^i = \emptyset$. Thus, we have:
\begin{align}
    \sum_{e \in \E: V(e) \cap U^i = \emptyset} w^i_e x^i_e \le \sum_{e \in \E: V(e) \cap U^i = \emptyset} w^i_e x_e.\label{IS Conditions Proof 2}
\end{align}
Combining \eqref{IS Conditions Proof 1} and \eqref{IS Conditions Proof 2} gives
\begin{align}
    \sum_{e \in \E: V(e) \cap U^i \ne \emptyset} w^i_e x^i_e  > \sum_{e \in \E: V(e) \cap U^i \ne \emptyset} w^i_e x_e. \label{IS Conditions Proof 4}
\end{align}
Finally, note that by construction of $U^i$, if for an exchange $e \in X^i$ it holds that $V(e) \cap U^i \ne \emptyset$, then actually $V(e) \subseteq U^i$. These exchanges thus describe a feasible KEP solution on $G[U^i]$. We now have
\begin{align}
    \beta(U^i) \ge \sum_{e \in \E: V(e) \cap U^i \ne \emptyset} w^i_e x^i_e. \label{IS Conditions Proof 5}
\end{align}
From combining \eqref{IS Conditions Proof 4} and \eqref{IS Conditions Proof 5}, we can conclude that \eqref{IS Conditions} is violated for the set $U^i$.
\end{proof}

We can now replace Constraints~\eqref{IP:MaxRP-KEP:agentAccept} by Constraints~\eqref{IS Conditions}, for each set $U^i \subseteq V^i$ for each agent $i \in \P$. We refer to these constraints as the {\em subset rejection constraints}. Violation of any such constraint means the agent has incentive to reject all exchanges making use of vertices in this subset and replacing it with a maximum agent value KEP solution. By this replacement, we obtain the following formulation.
\begin{subequations} \label{IP:RP-KEP-Reform}
\begin{align}
    \text{maximize} \ \sum_{e \in \E} w_e x_e  \label{IP:RP-KEP-Reform: Obj}\\
    \text{subject to} \ \sum_{e \in \E: v \in V(e)} x_e\ & \le\ 1 & \forall v \in V \\
   \sum_{e \in \E: V(e) \cap U^i \ne \emptyset} w^i_e x_e\ &  \ge\ \beta(U^i) & \forall i \in \P, U^i \subseteq V^i \label{IP:RP-KEP-Reform-RPcons}\\
   x_e\ & \in\ \{0,1\} & \forall e \in \E  \label{Model: Cycle Master Binary}
\end{align}
\end{subequations}
As $\beta(U^i)$ is independent of the value of other variables, this is a single-level reformulation of the problem. However, the formulation is exponential in size due to the number of subsets $U^i \subseteq V^i$. Due to $\Sigma_2^p$-completeness, we cannot hope for a subexponential size formulation unless the polynomial hierarchy collapses.

\subsection{Solving MaxRP-KEP} \label{sec:Solving}
In this section, we discuss our solution approach for MaxRP-KEP. The size of this model is exponential in the number of vertices of the graph, since for every agent $i$, it requires one constraint for each subset $U^i \subseteq V^i$. In our solution method, we start with a relaxed version of this model, dubbed MaxRP-KEP-R, using only a small collection of the subset rejection constraints~\eqref{IS Conditions}. This gives rise to an intuitive row generation algorithm, in which iteratively MaxRP-KEP-R is solved to optimality and violated subset rejection constraints are identified for each agent. We will discuss the identification of optimal solutions to MaxRP-KEP-R and the separation of violated subset rejection constraints in some more detail.


\subsubsection{Identifying optimal solutions to MaxRP-KEP-R.}\label{sec:Solving:Opt MaxRP-KEP-R}
It has been observed that kidney exchange problems usually have many optimal solutions~\citep{farnadi2021individual}. This carries over to MaxRP-KEP-R, in particular when the number of rejection constraints is still small, as the problem is then almost similar to a regular KEP. Selecting a solution among the different optima is a delicate matter, as some solutions may be rejection-proof, while others are not. It is desirable to identify a rejection-proof optimal solution if it exists, as then we can restrict the number of iterations of the (costly) procedure of separating violated subset rejection constraints and re-solving of the resulting, strengthened, MaxRP-KEP-R. The following lemma gives us insight in how to compare solutions in terms of rejection-proofness.

\begin{lemma}\label{lemma:compare rejection-proof}
Consider two solutions $X, Y \in \X$ such that
\begin{enumerate}[label=(\alph*)] 
    \item $(Y \cap \E^0) \subseteq (X \cap \E^0)$,\label{enum:compare1}
    \item $w^i(Y) \ge w^i(X)$ for all agents $i \in \P$.\label{enum:compare2}
\end{enumerate}
If $X$ is rejection-proof, then $Y$ is rejection-proof.
\end{lemma}
\begin{proof}
Suppose on the contrary that $Y$ is not rejection-proof. Then, for some agent $i \in \P$, there exists a solution $Y^i$ optimal to RKEP$(Y, i)$ such that $w^i(Y^i) > w^i(Y)$. Note that any solution feasible to RKEP$(Y,i)$ is also feasible to RKEP$(X,i)$ because of Condition~\ref{enum:compare1}. Finally, $w^i(Y^i) > w^i(Y) \ge w^i(X)$ means the agent has incentive to reject $X$ as well, meaning that $X$ is also not rejection-proof.
\end{proof}

This lemma shows that the number of shared exchanges in a solution and agent value are two factors influencing whether or not a solution is rejection-proof. Indeed, when the number of shared exchanges in a solution $X$ is small, this implies that the number of feasible rejection strategies, i.e. the number of feasible solutions of RKEP$(X,i)$, is more restricted compared to when there are many shared exchanges. This feasible region is also more restricted if the agent value $w^i(X)$ of $i$ for solution $X$ is larger, which is consistent with Constraints~\eqref{IP:MaxRP-KEP:agentAccept}.\\

We use this knowledge to propose as a tiebreaker the internal preference objective
\begin{equation}\label{eq:tiebreak:internal}
    t_1(X) := \sum_{i \in \P} \sum_{e \in \E^i} w^i_e x_e.
\end{equation}
By counting only the agent value of internal exchanges, it immediately takes into account both conditions of Lemma~\ref{lemma:compare rejection-proof}. Clearly, this tiebreaker prioritizes solutions with high agent values, whereas solutions with many shared exchanges are disadvantaged. In fact, we can show that given two solutions as in Lemma~\ref{lemma:compare rejection-proof}, the tiebreaking rule will always value the solution more likely to be rejection-proof more highly.

\begin{lemma}\label{lemma:tiebreak rule}
Consider two solutions, $X$ and $Y$, such that $(Y \cap \E^0) \subseteq (X \cap \E^0)$ and $w^i(Y) \ge w^i(X)$ for all agents $i \in \P$.
Then, we have $t_1(Y) \ge t_1(X)$, i.e., the tiebreak value of $Y$ is at least as high as the tiebreak value of $X$.
\end{lemma}
\begin{proof}
Let $(Y \cap \E^0) \subseteq (X \cap \E^0)$. Then, using the definitions of $w^i(Y)$ and $w^i(X)$, we get
\begin{align}\label{eq:tiebreak1}
    \sum_{e \in \E^0} w^i_e y_e \le \sum_{e \in \E^0} w^i_e x_e \Rightarrow \sum_{e \in \E^i} w^i_e y_e \ge \sum_{e \in \E^i} w^i_e x_e.
\end{align}
Now, it follows from~\eqref{eq:tiebreak:internal} and~\eqref{eq:tiebreak1} that $t_1(Y) \ge t_1(X)$.
\end{proof}

Notice that there is a strong connection between this tiebreaking objective and solutions obtained from the MaxInt mechanism. In fact, the same two objectives are considered, namely maximizing total number of transplants and maximizing the number of internal transplants, but in a different order. As MaxInt proposes rejection-proof KEP solutions, this supports the use of this tiebreaker.

The tiebreaker is implemented as follows. Let $Z$ be an upper bound to the agent value of internal exchanges in any solution ($Z \ge \max_{X \in \X} \sum_{i \in \P} \sum_{e \in \E^i} w^i_e x_e$). We then replace the objective~\eqref{IP:MaxRP-KEP:obj} of MaxRP-KEP, by
\begin{align}
    \sum_{e \in \E} Z w_e x_e + \sum_{i \in \P} \sum_{e \in \E^i} w^i_e x_e. \label{Obj:IntPref}
\end{align}
This objective takes into account both the social value of all transplants $(\sum_{e \in \E} Zw_e x_e)$ and the agent value of internal exchanges $(\sum_{e \in \E^i} w^i_e x_e)$. Due to the heavy weighting of social value, the agent value will only function as a tiebreaker between solutions with equal social value.

\subsubsection{Separating violated subset rejection constraints}\label{Sec:Solving:Violated}
Consider a non-rejection-proof solution $X$. In the proof of Theorem~\ref{Theorem: Stability cycle Form}, we show that if there exists an agent $i \in \P$ and solution $X^i$ feasible to RKEP$(X, i)$ with $w^i(X^i) > w^i(X)$, at least one subset rejection constraint must be violated. In the proof, we showed the constraint for agent $i$ and subset
\begin{align*}
    U^i = \{v \in V^i \mid \exists e \in \E: v \in V(e), x_e^i = 1, x_e = 0 \}.
\end{align*}
is then violated.

The right-hand side of the violated rejection constraint can be computed as follows.
\begin{align}
  \beta(U^i) = \sum_{e \in \E: V(e) \subseteq U^i} w^i_ex^i_e.
\end{align}

\section{Maximum Weight Internal Matching} \label{sec:MaxIntMech}
In this section, we describe a first and simple rejection-proof mechanism, dubbed the \textit{Maximum Weight Internal KEP} mechanism, shortened to the MaxInt mechanism or simply MaxInt. This mechanism proposes a solution where the total transplant value is maximized, under the restriction that the value of internal transplants is maximum.

\begin{definition}{Maximum Weight Internal Transplants (MaxInt)}\\
The MaxInt mechanism selects a KEP solution $X \in \X$ optimizing the following integer program (MaxInt-KEP).
\end{definition}
\begin{subequations} \label{IP:maxint}
\begin{align}
    \text{maximize} \quad \sum_{e \in \E} w_e x_e  \label{IP:maxint:Objective}\\
    \text{subject to} \quad \sum_{e \in \E: v \in V(e)} x_e \ &\le \ 1 &\forall v \in V \label{IP:maxint:vertex}\\
    \sum_{e \in \E^i} w^i_e x_e \ &= \ \beta(V^i) & \forall i \in \P \label{IP:maxint:eq} \\
    x_e \ &\in \ \{0,1\} &\forall e \in \E \label{IP:maxint:binary}
\end{align}
\end{subequations}

Constraints~\eqref{IP:maxint:vertex} again impose that $X$ induces a set of pairwise vertex-disjoint exchanges. Constraints~\eqref{IP:maxint:eq} ensure that for each individual agent $i \in \P$, the exchanges of $X$ induced by $V^i$ forms a maximum weight internal KEP solution. Notice that these constraints are satisfied whenever every agent selects a greedy withholding strategy. Nevertheless, an agent might have many distinct greedy withholding strategies, each of which might have a different impact on the number of transplants that can be realized overall. Therefore, MaxInt can thus be seen as mechanism that coordinates a profile of greedy withholding strategies allowing the maximum overall value of transplants in the shared kidney exchange pool.


\begin{theorem}
The MaxInt mechanism is rejection-proof.
\end{theorem}
\begin{proof}
Let $X \in \X$ be a KEP solution proposed by MaxInt, encoded by binary variables $(x_e)_{e \in \E}$, and $i \in \P$ an arbitrary agent. Then
\begin{equation*}
    w^i(X) = \sum_{e \in \E^i} w^i_e x_e + \sum_{e \in \E^0} w^i_e x_e.
\end{equation*}
For any optimal solution $X^i$ to RKEP$(X, i)$, encoded by binary variables $(x^i_e)_{e \in \E}$, we have
\begin{equation*}
    \text{RKEP}(X, i) = \sum_{e \in \E^i} w^i_e x^i_e + \sum_{e \in \E^0} w^i_e x^i_e.
\end{equation*}
As $X \cap \E^i$ is a maximum weight internal set of exchanges on $V^i$, we have
\begin{equation}\label{eq:maxint_ub_internal}
\sum_{e \in \E^i} w^i_e x_e \ge \sum_{e \in \E^i} w^i_e x^i_e.
\end{equation}
Next, notice that a shared exchange $e$ can only be contained in $X^i$ if it was in the proposed solution $X$ in the first place, hence
\begin{equation}\label{eq:maxint_ub_shared}
    \sum_{e \in \E^0} w^i_e x_e \ge \sum_{e \in \E^0} w^i_e x^i_e.
\end{equation}
It follows for adding~\eqref{eq:maxint_ub_internal} and~\eqref{eq:maxint_ub_shared} that $w^i(X) \ge \text{RKEP}(X, i)$, showing MaxInt is rejection-proof.
\end{proof}

\section{Computational Results} \label{sec:CompResults}
This section gives the outcomes of both rejection-proof mechanisms, with regards to total transplants. We consider the same instances and experimental setting as discussed in Section  \ref{sec:WithvsReject}. We are now in a position to evaluate the outcomes of the MaxInt and MaxRP-KEP mechanisms, as described in Sections \ref{sec:RPKEP} and \ref{sec:MaxIntMech}. First, we will compare the number of transplants resulting from the mechanisms against the social optimum, assuming agents cooperate fully. Second, we will look at the incentives for agents. Clearly, when rejection-proof mechanisms are employed, rejection strategies provide no benefit to agents. We will therefore only check whether withholding strategies still provide a benefit to agents.

\begin{table}[htb]
    \centering\footnotesize
    \begin{tabular}{ccccccccc}
        \toprule

        & & \multicolumn{3}{c}{Delorme SplitPRA-BXM-PRA0} & & \multicolumn{3}{c}{Saidman}\\

        \midrule

        Instance set & & MaxRP-KEP & MaxInt & All Withhold & &  MaxRP-KEP & MaxInt & All Withhold \\

        \midrule
        10x10 & & 0.994 & 0.944 & 0.943 & & 0.999 & 0.885 & 0.868 \\
        10x12 & & 0.996 & 0.934 & 0.931 & & 1.000 & 0.911 & 0.890 \\
        10x14 & & 0.996 & 0.944 & 0.944 & & 1.000 & 0.922 & 0.900 \\
        10x16 & & 0.997 & 0.940 & 0.938 & & 0.999 & 0.920 & 0.899 \\
        10x2 & & 1.000 & 0.935 & 0.935 & & 0.998 & 0.891 & 0.855 \\
        10x4 & & 0.991 & 0.882 & 0.882 & & 0.994 & 0.866 & 0.846 \\
        10x6 & & 0.996 & 0.931 & 0.926 & & 0.996 & 0.868 & 0.844 \\
        10x8 & & 0.994 & 0.928 & 0.927 & & 0.995 & 0.885 & 0.869 \\
        15x10 & & 0.993 & 0.908 & 0.905 & & 1.000 & 0.910 & 0.886 \\
        15x2 & & 0.996 & 0.947 & 0.939 & & 0.996 & 0.888 & 0.862 \\
        15x4 & & 0.990 & 0.879 & 0.873 & & 0.996 & 0.871 & 0.846 \\
        15x6 & & 0.996 & 0.892 & 0.877 & & 0.998 & 0.888 & 0.860 \\
        15x8 & & 0.996 & 0.910 & 0.907 & & 0.999 & 0.891 & 0.867 \\
        30x2 & & 0.992 & 0.885 & 0.854 & & 0.997 & 0.915 & 0.901 \\
        30x4 & & 0.990 & 0.843 & 0.813 & & 0.999 & 0.896 & 0.871 \\
        30x6 & & 0.995 & 0.866 & 0.839 & & 1.000 & 0.912 & 0.881 \\
        50x2 & & 0.992 & 0.858 & 0.820 & & 0.998 & 0.940 & 0.928 \\
        80x2 & & 0.992 & 0.845 & 0.807 & & 1.000 & 0.959 & 0.958 \\
        \bottomrule
    \end{tabular}
    \caption{The columns ``MaxRP-KEP'' and ``MaxInt'' show the average fraction of the social optimum (maximum number of transplants provided no rejection / withholding strategies are used) achieved by MaxRP-KEP and MaxInt respectively, over all 50 instances per Delorme and Saidman instance set. The column ``All Withhold shows'' the average fraction of the social optimum whenever each agent uses the greedy withholding strategy.\label{tab:mechanismloss}}
\end{table}

Table~\ref{tab:mechanismloss} shows that the MaxRP-KEP mechanism achieves rejection-proofness at a small cost on average. For all instance sets, the number of transplants lost, compared to the social optimum is at most 1\%. These losses are consistently less than those suffered if only a single agent employed either rejection or withholding strategies (Table~\ref{tab:agentStrat}). The MaxInt mechanism comes at a higher cost, with losses up to 15\% of total transplants. This is still an improvement over the losses suffered in case all agents employ greedy withholding strategies, which are shown in the final column of the table. Finally, Table~\ref{tab:stratwithmech} in the appendix shows the outcome agents can achieve using greedy withholding strategies if the rejection-proof mechanisms are employed. Contrary to the situation where a maximum weight solution on the common pool is selected (Table~\ref{tab:agentStrat}), agents generally lose transplants by withholding. This is especially the case for agents with large pools, where average losses can be up 8 \% for the MaxRP-KEP mechanism.

\section{Discussion} \label{Sec:Disc}
In this paper, we address the issue of cooperation in kidney exchange from a new angle. Previous work focused on incentivizing agents to contribute all of their patients and donors to a common pool. The underlying assumption of this work is that agents will strategically choose to reveal or withhold their pairs from a cooperation, so as to maximize the value of transplants to their patients (withholding strategies). One of our contributions is to recognize a second option for agents, the rejection strategies. In Section \ref{sec:WithvsReject}, we showed that such strategies provide better outcomes for the agent than existing withholding strategies with reasonable computational and information requirements. Furthermore, we showed such rejection strategies cause an overall loss of transplants. As a result, we propose that rejection-proofness is a valuable property for cooperation mechanisms.

We provide two different rejection-proof mechanisms, MaxInt and MaxRP-KEP. MaxInt is simple to implement and easy to understand, which may carry benefits in practice. Currently, international cooperations exist~\citep{biro2020} using a \textit{consecutive pool} concept. Countries first match internally and send all unmatched pairs to a common pool. This corresponds to agreeing that every country will use the greedy withholding strategy. The MaxInt mechanism is very similar in spirit, but coordinates the greedy withholding strategies so as to maximize the total number of transplants. The second mechanism, MaxRP-KEP, can provide significantly higher numbers of transplants. This performance does come at a cost with regards to implementation and computational difficulty. Further research into algorithms for solving this problem is required to allow for cooperations with many agents with large pools.

Common to both mechanisms is that besides their designed outcome of nullifying rejection strategies, they also both eliminate expected benefits provided by greedy withholding strategies. We do note that they are not strategy-proof, as evidenced by the existence of instances where the greedy withholding strategy does provide a benefit.

As a final remark about the benefits of our mechanisms, we note that they provide a natural way of reconciling differences in transplant valuation in cooperations. These mechanisms make use of agent $(w^i_e)$ and cooperation $(w_e)$ valuations. Through the constraints ensuring rejection-proofness, agents are assured that their private valuations will be taken into account. This allows participants in a cooperation to set their own priorities.

Looking forward, we believe combinations of rejection-proof and credit mechanisms are interesting to study. Current credit mechanisms do not handle failure or rejection of proposed transplants. By removing incentives to reject this issue is ameliorated. Rejection-proof mechanisms provide guarantees that agents will not lose out by cooperating, but can still allow for unequal division of the benefits of cooperation. Furthermore, in our experiments rejection-proof mechanisms still allow situations where greedy withholding strategies are profitable. Depending on graph structure, there may be situations where they are consistently profitable. Credit mechanisms can reduce these issues in multi-period settings. By combining rejection-proof and credit mechanisms, the risk of strategic behaviour can be further reduced.
\section*{Acknowledgements}
The authors thank P\'eter B\'ir\'o for helpful discussions. The research of Frits Spieksma is supported by Dutch Research Council (NWO) Gravitation
Project NETWORKS [Grant 024.002.003].

\bibliographystyle{plainnat}
\bibliography{KEP}

\newpage

\setcounter{table}{0}
\renewcommand{\thetable}{\Alph{section}.\arabic{table}}
\renewcommand{\thefigure}{\Alph{section}.\arabic{figure}}
\begin{appendices}
\section{Supporting tables}

\begin{table}[htbp]
    \centering\footnotesize
    \begin{tabular}{lrrrr}
        \toprule
        Instance Set &  Total Time (s) &  Constraints Added &  Iterations &  Instances Solved \\
        \midrule
        DelBXMPRA0Score\_96\_4\_1l2m   &            0.46 &               5.50 &        4.77 &                30 \\
        DelBXMPRA0Score\_96\_4\_1l5s   &            0.17 &               3.37 &        3.80 &                30 \\
        DelBXMPRA0Score\_96\_4\_2equal &            1.15 &               9.67 &        7.63 &                30 \\
        DelBXMPRA0Score\_96\_4\_4equal &            0.06 &               3.13 &        2.70 &                30 \\
        DelBXMPRA0\_96\_4\_1l2m        &            0.09 &               1.60 &        2.47 &                30 \\
        DelBXMPRA0\_96\_4\_1l5s        &            0.08 &               1.07 &        1.93 &                30 \\
        DelBXMPRA0\_96\_4\_2equal      &            0.09 &               1.40 &        2.17 &                30 \\
        DelBXMPRA0\_96\_4\_4equal      &            0.04 &               1.20 &        1.83 &                30 \\
        SaidmanScore\_96\_4\_1l2m      &          308.04 &              36.42 &       22.83 &                24 \\
        SaidmanScore\_96\_4\_1l5s      &          303.07 &              33.26 &       28.67 &                27 \\
        SaidmanScore\_96\_4\_2equal    &          732.42 &              41.83 &       28.04 &                24 \\
        SaidmanScore\_96\_4\_4equal    &          518.13 &              26.64 &       13.89 &                28 \\
        Saidman\_96\_4\_1l2m           &            2.97 &               3.40 &        3.37 &                30 \\
        Saidman\_96\_4\_1l5s           &            5.75 &               7.80 &        7.40 &                30 \\
        Saidman\_96\_4\_2equal         &            4.70 &               2.50 &        2.77 &                30 \\
        Saidman\_96\_4\_4equal         &            2.50 &               9.07 &        6.87 &                30 \\
        \bottomrule
    \end{tabular}
    \caption{Comparison of computational results for all instance sets. All results were obtained using no initial constraints, the weighted tiebreaking and maximum violation constraints added. Instances solved is the amount of instances solved to optimality in less than 3 hours (10 800s), out of 30. Computation time, constraints added and number of iterations are the means of instances solved to optimality within this time.}
    \label{Table: Instance Sets}
\end{table}

\begin{table}[htb]
    \centering\footnotesize
    \begin{tabular}{ccccccc}
        \toprule
        & & \multicolumn{2}{c}{Delorme SplitPRA-BXM-PRA0} & & \multicolumn{2}{c}{Saidman}\\
         \midrule
        Instance Set & & WA MaxRP-KEP & WA MaxInt & & WA MaxRP-KEP & WA MaxInt \\
        \midrule
        10x10 & & 1.005 & 1.000 & & 1.009 & 0.987\\
        10x12 & & 1.000 & 1.000 & & 1.000 & 0.979 \\
        10x14 & & 1.008 & 1.004 & & 1.008 & 0.985 \\
        10x16 & & 1.020 & 1.000 & & 0.997 & 0.991 \\
        10x2  & & 1.000 & 1.000 & & 0.996 & 0.985 \\
        10x4  & & 0.991 & 1.011 & & 0.987 & 0.996 \\
        10x6  & & 1.007 & 1.000 & & 0.976 & 1.004 \\
        10x8  & & 1.006 & 1.000 & & 0.991 & 1.000 \\
        15x10 & & 1.008 & 1.000 & & 0.989 & 0.976 \\
        15x2  & & 1.000 & 1.000 & & 0.972 & 0.990 \\
        15x4  & & 0.986 & 1.005 & & 0.998 & 0.993 \\
        15x6  & & 0.996 & 0.980 & & 1.002 & 0.976 \\
        15x8  & & 0.994 & 0.994 & & 0.979 & 0.993 \\
        30x2  & & 0.970 & 0.976 & & 0.984 & 0.991 \\
        30x4  & & 0.964 & 0.992 & & 0.985 & 0.987 \\
        30x6  & & 0.988 & 0.982 & & 0.987 & 0.983 \\
        50x2  & & 0.945 & 0.974 & & 0.981 & 0.991 \\
        80x2  & & 0.917 & 0.971 & & 0.994 & 1.000 \\
        \bottomrule
    \end{tabular}
    \caption{This table shows the number of transplants achieved by an agent employing the greedy withholding strategy (WA: Withhold Agent), divided by the number of transplants without withholding, for both rejection-proof mechanisms and over all 50 instances in the set.\label{tab:stratwithmech}}
\end{table}

\clearpage
\section{Supporting figures}
\setcounter{figure}{0}

\begin{figure}[htb!]
       \centering
       \begin{subfigure}[b]{0.49\linewidth}
            \centering
             \includegraphics[scale = 0.55]{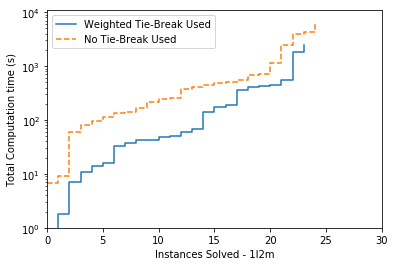}
            \caption{\label{fig:tiebreak:noinit:maxviol:1l2m}}
        \end{subfigure}
        \hfill
        \begin{subfigure}[b]{0.49\linewidth}
            \centering
            \includegraphics[scale = 0.55]{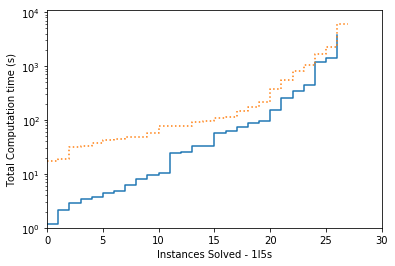}
            \caption{\label{fig:tiebreak:noinit:maxviol:1l5s}}
        \end{subfigure}
        \begin{subfigure}[b]{0.49\linewidth}
            \centering
            \includegraphics[scale = 0.55]{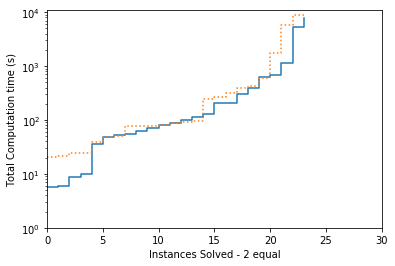}
            \caption{\label{fig:tiebreak:noinit:maxviol:2equal}}
        \end{subfigure}
        \begin{subfigure}[b]{0.49\linewidth}
            \centering
            \includegraphics[scale = 0.55]{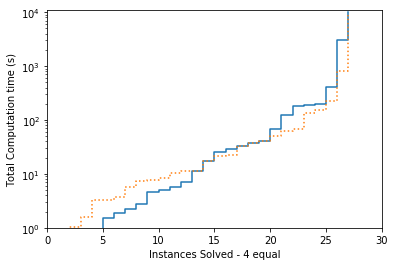}
            \caption{\label{fig:tiebreak:noinit:maxviol:4equal}}
        \end{subfigure}
        \caption{Comparisons of solution times dependant on use of the tiebreak rule. \label{Fig:tiebreak}}
\end{figure}
\clearpage
\section{Proof of Theorem~\ref{th:complexity_proof}}
\setcounter{figure}{0}
This theorem considers the computational complexity of the optimization problem faced by the MaxRP-KEP mechanism. We focus ourselves on the case of unit transplant values, i.e. $w_e = |V(e) \cap \left(\bigcup_{i \in \P} P^i\right)|$ and $w_e^i = |V(e) \cap P^i|$ for all exchanges $e \in \E$. We denote by DecRP-KEP the decision variant of this problem, which is described as follows.

\subsection*{DecRP-KEP}
{\bf Instance:} a compatibility graph $G = (V,A)$ and a set of agents $\P$, such that $V = \cup_{i \in \P} V_i$ consists of the pools $V^i = P^i \cup N^i$ of donor-recipient pairs and non-directed donors for agent $i \in \P$, and $A$ the set of potential transplants. Furthermore, we are given a maximum cycle length $\cyclen$, maximum chain length $\chainlen$, and a nonnegative integer $t$.\\
{\bf Question:} does there exist a rejection-proof KEP solution consisting of vertex-disjoint cycles of length at most $\cyclen$ and chains of length at most $\chainlen$, covering at least $t$ vertices?\\[1ex]

Theorem~\ref{th:complexity_proof} claims that for $K \ge 3$ fixed, and any $L \in \N$, DecRP-KEP belongs to the class $\Sigma_2^p$, and furthermore that it is a complete with respect to this class. We prove this statement by providing a reduction from Adversarial (2,2)-SAT, which is proven to be $\Sigma_2^p$-complete in~\cite{Smeulders2022Stackel}. This problem is closely related to the foundational $\Sigma_2^p$-complete problem, $B_2^{CNF}$.\newline\newline

\noindent {\bf Problem:} Adversarial (2,2)-SAT\\
{\bf Instance:} Boolean CNF formula $\varphi$ consisting of two sets $X, Y$ of variables, and a set $C$ of disjunctive clauses over $X$ and $Y$, such that each variable $z \in X \cup Y$ occurs twice in negated ($\neg z$), twice in unnegated form ($z$) in $C$.\\
{\bf Question:} does there exist a truth setting $\theta_X \colon X \to \{\true, \false\}$ such that for any truth setting $\theta_Y \colon Y \to \{\true, \false\}$, the formula $\varphi$ is not satisfied.\newline\newline

We will say that a variable $z \in X \cup Y$ occurs in sense $\psi = t$ or $\psi = f$ in clause $c \in C$, if $c$ contains the literal $z$ or $\neg z$ respectively. Furthermore, we denote by $c(z, i)$ and $c(\neg z, i)$ the clause in which $z$ or $\neg z$ respectively occur for the $i$-th time, where $i \in \{1,2\}$. In the following sections, we prove that polynomial reducibility of Adversarial (2,2)-SAT to DecRP-KEP. First, we describe a mapping $f$ of an Adversarial (2,2)-SAT instance $\varphi$ to an instance $f(\varphi)$ of DecRP-KEP. Next, we lay out specific properties of optimal and feasible solutions of the instances $f(\varphi)$. Finally, we bring all of this together in the polynomial reduction.

\subsection*{Instance mapping}
Let us be given an instance $\varphi = \varphi(X,Y,C)$ to Adversarial (2,2)-SAT. We construct an instance $f(\varphi) = (G, \P, K, L, t)$ to DecRP-KEP forming a 2-agent kidney exchange program, i.e., $\P = \{b,g\}$. Informally, we will refer to these agents as the blue and the green agent. The figures of the instance use these same colours to differentiate vertices of the different agents.\\[1ex]
{\bfseries $X$-gadgets:} for each variable $x \in X$, we define the vertex set $V_x \coloneqq V_x^g \cup V_x^b$, with agent's vertex sets $V^g_x$ and $V^b_x$ of agent $g$ and $b$ respectively, as follows.
    \[V^g_x \coloneqq \{x_0\} \cup \Big\{x_{\psi, i} : \psi \in \{t,f\}, i \in \{1,2\}\Big\} \]
    and
    \[V^b_x \coloneqq \Big\{\gamma_{x,\psi, i} : \psi \in \{t,f\}, i \in \{1,2\}\Big\} \cup \Big\{\alpha_{x, i} : i \in \{1,2\} \Big\}\]
    The arc set $A_x \subseteq \{(i,j) : i,j \in V_x\}$ is defined as
    \begin{align*}
        A_x \coloneqq &\Big\{(x_0, x_{\psi,1}), (x_{\psi,1}, x_{\psi,2}), (x_{\psi,2},x_0) : \psi \in \{t,f\}\Big\}\\
                &\cup \Big\{(x_{\psi, i}, \alpha_{x, i}), (\alpha_{x, i}, \gamma_{x, \psi, i}), (\gamma_{x, \psi, i}, x_{\psi, i}) : \psi \in \{t,f\}, i \in \{1,2\}\Big\}
    \end{align*}

\begin{figure}[htb!]
    \centering
    \begin{subfigure}[b]{0.3\linewidth}
        \centering
        \includegraphics[scale = 0.7]{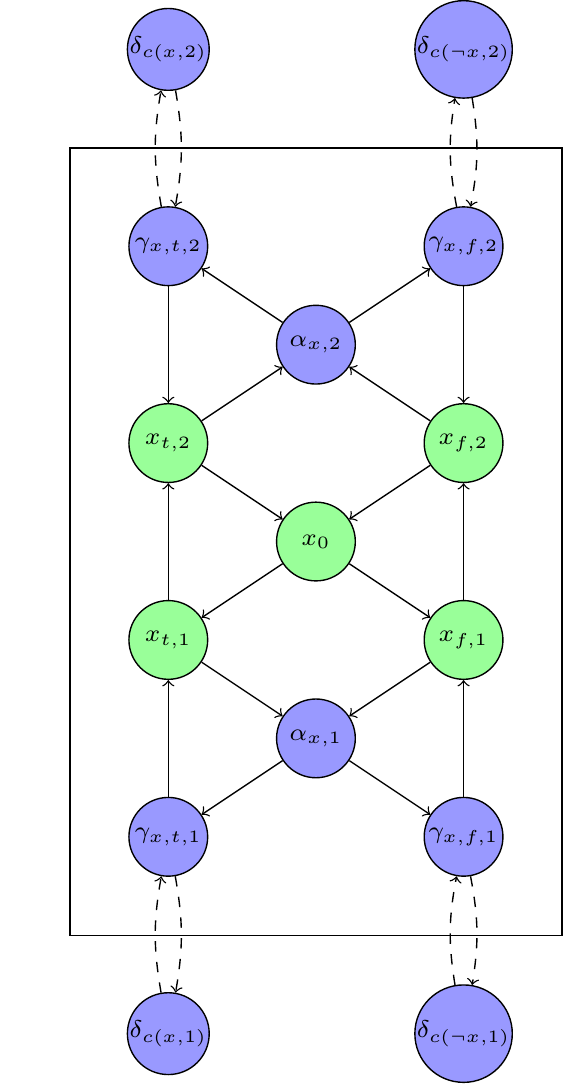}
        \caption{\label{fig:gadget_xvar}Gadget for variable $x \in X$}
    \end{subfigure}
    \begin{subfigure}[b]{0.3\linewidth}
        \centering
        \includegraphics[scale = 0.7]{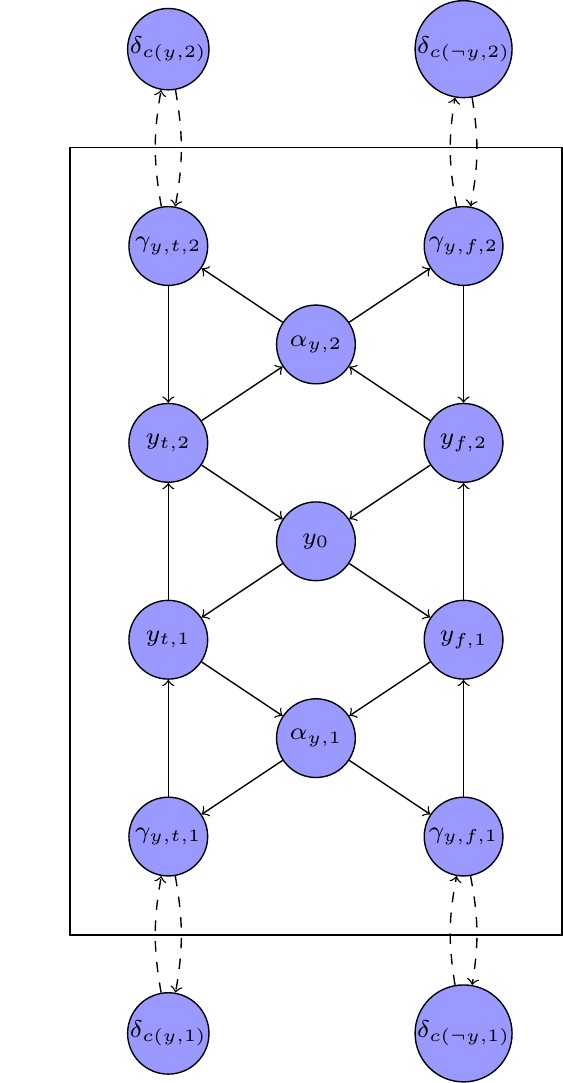}
        \caption{\label{fig:gadget_yvar}Gadget for variable $y \in Y$}
    \end{subfigure}
    \begin{subfigure}[b]{0.3\linewidth}
        \centering
        \includegraphics[scale = 0.7]{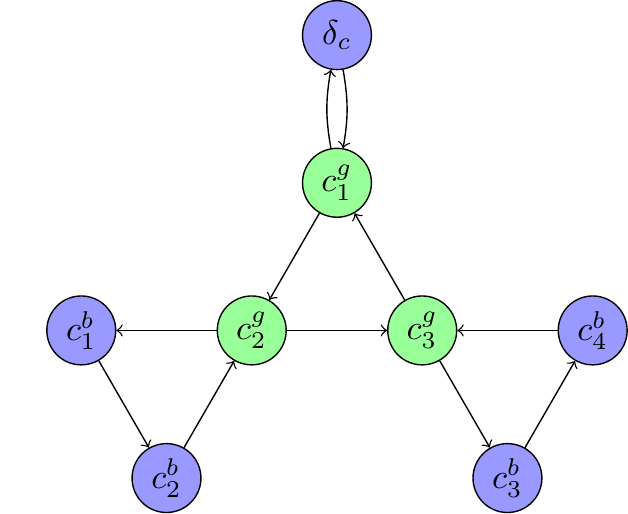}
        \caption{\label{fig:gadget_sat}Gadget for clause $c \in C$}
    \end{subfigure}
    \caption{Gadgets for the polynomial reduction defining the transformed instance $f(\varphi)$.\label{fig:gadgets}}
\end{figure}

    The graph $G_x = (V_x, A_x)$ will be referred to as the {\em $x$-gadget} or {\em variable gadget} of $x$, and is depicted within the black square in Figure~\ref{fig:gadget_xvar}.\\[1ex]
{\bfseries $Y$-gadgets:} for each variable $y \in Y$, we obtain the vertex set $V_y \coloneqq V_y^g \cup V_y^b$ of agent $b$ and $g$ respectively. Now, $V^g_y = \emptyset$ and
    \begin{align*}
        V^b_y \coloneqq &\{y_0\} \cup \Big\{y_{\psi,i} : \psi \in \{t,f\}, i \in \{1,2\}\Big\}\\
                 &\cup \Big\{\gamma_{y, \psi, i} : \psi \in \{t,f\}, i \in \{1,2\}\Big\} \cup \Big\{\alpha_{y,i} : i \in \{1,2\} \Big\}
    \end{align*}
    Similar to the arc set $A_x$ for each $x$-gadget, the arc set $A_y \subseteq \{(i,j) : i,j
    \in V_y\}$ is defined as
    \begin{align*}
        A_y \coloneqq &\Big\{(y_0, y_{\psi,1}), (y_{\psi,1}, y_{\psi,2}), (y_{\psi,2},y_0) : \psi \in \{t,f\}\Big\}\\
                &\cup \Big\{(y_{\psi,i}, \alpha_{y,i}), (\alpha_{y,i}, \gamma_{y, \psi, i}), (\gamma_{y, \psi, i}, y_{\psi, i}) : \psi \in \{t,f\}, i \in \{1,2\}\Big\}
    \end{align*}
    The graph $G_y = (V_y, A_y)$ will be referred to as the {\em $y$-gadget} or {\em variable gadget} of $y$, and is illustrated within the black square in Figure~\ref{fig:gadget_yvar}.\\[1ex]
    {\bfseries Clause gadgets:} for each clause $c \in C$, we define the vertex set $W_c \coloneqq W^g_c \cup W^b_c$, where the vertex sets $W^g_c$ and $W^b_c$ are respectively defined by
    \[W^g_c \coloneqq \Big\{c_i^g : i \in \{1,2,3\}\Big\},\]
    and
    \[W^b_c \coloneqq \{\delta_c\} \cup \Big\{c_i^b : i \in \{1,2,3,4\}\Big\}.\]
    The arc set $A_c$ within the $c$-gadget is defined as
    \begin{align*}
        A_c \coloneqq &\Big\{(c_1^g, c_2^g), (c_2^g, c_3^g), (c_3^g, c_1^g)\Big\} \cup \Big\{(\delta_c, c_1^g), (c_1^g, \delta_c)\Big\}\\
               &\Big\{(c_2^g, c_1^b), (c_1^b, c_2^b), (c_2^b, c_2^g)\Big\} \cup \Big\{(c_3^g, c_3^b), (c_3^b, c_4^b), (c_4^b, c_3^g)\Big\}
    \end{align*}
    The graph $G_c = (W_c, A_c)$ will be referred to as the {\em $c$-gadget} or the {\em clause gadget} of $c$.
    The $c$-gadget is illustrated in Figure~\ref{fig:gadget_sat}.\\[1ex]
{\bfseries Arcs connecting variable and clause gadgets:} finally, we also add arcs connecting $x$- and $y$-gadgets to $c$-gadgets. In particular, we define these arcs as
\begin{align*}
    A_{\gamma,\delta} &= \Big\{(\gamma_{z, t, i}, \delta_{c(z, i)}), (\delta_{c(z,i)}, \gamma_{z, t, i}) : z \in X \cup Y, i \in \{1,2\}\Big\}\\
     &\cup \Big\{(\gamma_{z, f, i}, \delta_{c(\neg z, i)}), (\delta_{c(\neg z,i)}, \gamma_{z, f, i}) : z \in X \cup Y, i \in \{1,2\}\Big\}
\end{align*}
These arcs are depicted in Figures~\ref{fig:gadget_xvar} and~\ref{fig:gadget_yvar} as dashed arcs. The arcs $A_{\gamma, \delta}$ induce a set of 2-cycles, which we will refer to as \emph{$(\gamma, \delta)$-cycles}. Summarizing the above, the instance $f(\varphi)$ to MaxRP-KEP consists of a graph $G = (V,A)$ with
\[V = (\bigcup_{x \in X} V_x) \cup (\bigcup_{y \in Y} V_y) \cup (\bigcup_{c \in C} W_c),\]
and
\[A = (\bigcup_{x \in X} A_x) \cup (\bigcup_{y \in Y} A_y) \cup (\bigcup_{c \in C} A_c) \cup A_{\gamma, \delta}.\]

Furthermore, we set the parameters $K = 3$, $L\in \N$ arbitrarily, and $t = 9|X|+ 9|Y|+ 5|C|+ 3$. We proceed by providing the relevant cycle and chain packings, i.e. feasible KEP solutions, for our hardness reduction. Given this characterization of the DecRP-KEP instance $f(\varphi)$, we will now identify some structured packings related to the gadgets and prove that these structures need to appear in optimal solutions.

Notice that we show the result here for maximum cycle length $K = 3$, but all arguments used in the proof below will stay in place for all fixed $K \ge 4$ through the following adaptation of the variable gadgets (for all $x \in X$ and $y \in Y$): subdivide each arc $(\gamma_{z, \psi, i}, z_{\psi, i})$ for $\psi \in \{t, f\}, i \in \{1,2\}$ into $K - 2$ arcs, where all intermediate nodes belong to the vertex set $V_z^b$ of the blue agent.

\subsection*{Relevant gadget packings}
Consider an $x$-gadget, and let $\theta_X \colon X \to \{\true, \false\}$ be a truth setting for $X$. We refer to the packing defined by
\[\begin{cases}
        S_{x,t} = \Big\{\{x_{t,1}, x_{t,2},x_0\}, \{\gamma_{x,f,1}, x_{f,1}, \alpha_{x,1}\}, \{\gamma_{x,f,2}, x_{f,2}, \alpha_{x,2}\}\Big\}, \ \text{ if } \theta_X(x) = \true,\\
        S_{x,f} = \Big\{\{x_{f,1},x_{f,2},\gamma_x\}, \{\gamma_{x,t,1}, x_{t,1}, \alpha_{x,1}\}, \{\gamma_{x,t,2}, x_{t,2}, \alpha_{x,2}\}\Big\}, \ \text{ if } \theta_X(x) = \false.
  \end{cases}
\]
as being {\it consistent} with setting $x$ to $\true$ or $\false$ respectively. We can define similar notions of $S_{y,t}$ and $S_{y,f}$ for $y$-gadgets for $y \in Y$, through replacing $X$ with $Y$ and $x$ with $y$.

\begin{figure}[htb]
        \centering
        \begin{subfigure}[b]{0.49\linewidth}
            \centering
             \includegraphics[scale = 0.8]{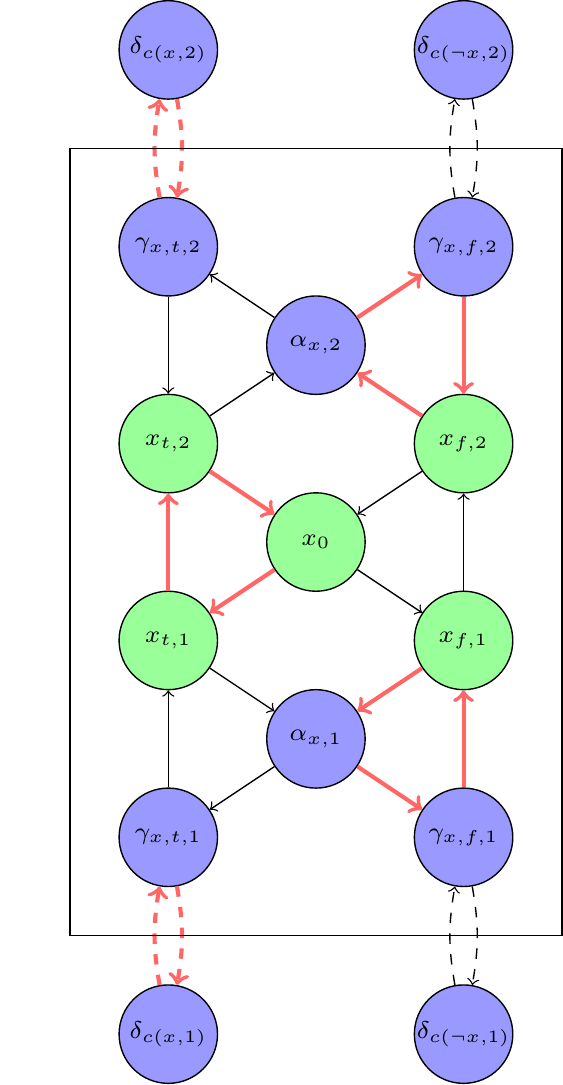}
            \caption{\label{fig:solution_xgadget_true}}
        \end{subfigure}
        \hfill
        \begin{subfigure}[b]{0.49\linewidth}
            \centering
            \includegraphics[scale = 0.8]{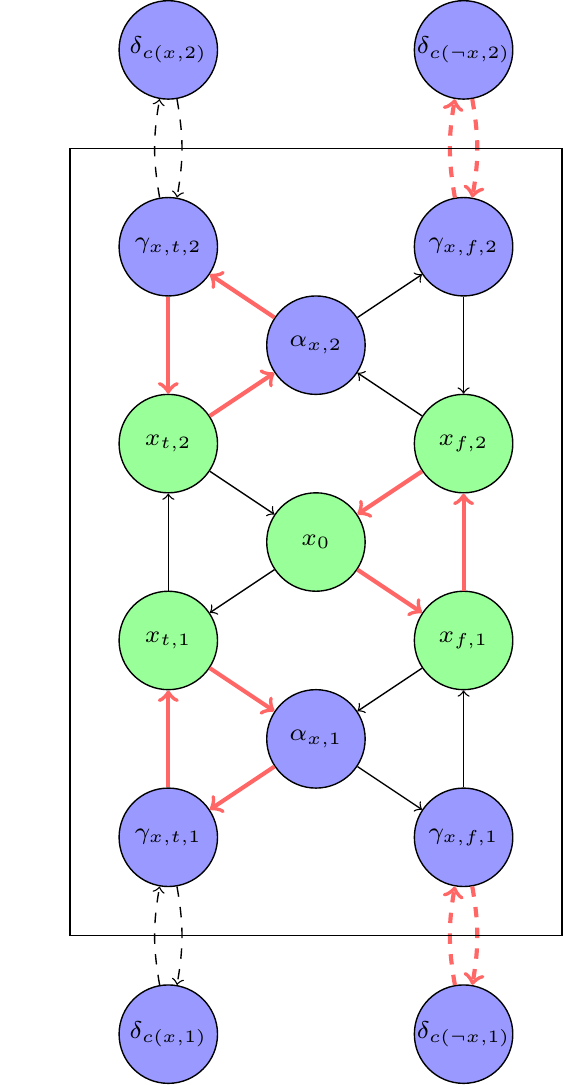}
            \caption{\label{fig:solution_xgadget_false}}
        \end{subfigure}
        \caption{Gadget packings for variable $x \in X$ consistent with \true~(\ref{fig:solution_xgadget_true}) and \false~(\ref{fig:solution_xgadget_false}). Note that the $(\gamma, \delta)$ cycles are not required for a packing to be consistent. These are coloured red to highlight the possibility of including them in a packing, if the $\delta$-vertices are not covered by another cycle already. Consistent packings work similarly for variables $y \in Y$, where the gadget consists of blue vertices only.\label{fig:solution_xgadget}}

    \end{figure}

\begin{figure}[htb]
    \centering
    \includegraphics[scale = 0.8]{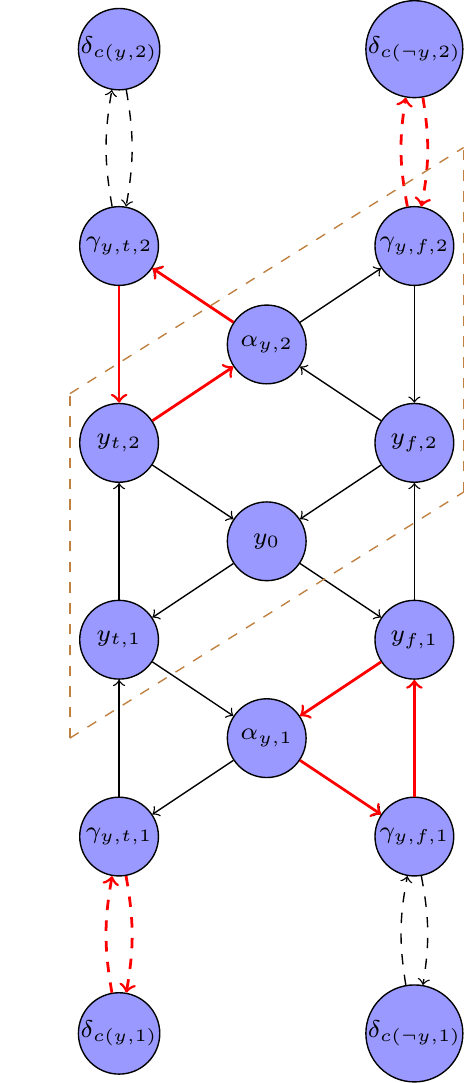}
    \caption{When a $y$-gadget has packing $S_3$, the vertices inside the brown parallelogram form a subset $U^b$ with a violated subset rejection constraint.\label{fig:solution_ygadget_zigzag}}
\end{figure}

Figure~\ref{fig:solution_xgadget} illustrates the gadget packings consistent with $\true$ and $\false$ respectively.
We show that for each rejection-proof solution of maximum size, i.e., a solution covering the largest number of vertices, we need to pack each variable gadget consistent with a truth setting.
\begin{lemma}\label{lemma:consistent_var_pack}
For each maximum solution to \emph{MaxRP-KEP} on $f(\varphi)$, each variable gadget (i.e., for all $x \in X$ and all $y \in Y$) is packed consistent with either \true~or \false.
\end{lemma}
\begin{proof}
We distinguish the proof for $x$- and $y$-gadgets, since the $x$-gadget contains pairs from the green agent, whereas the $y$-gadget does not.\\[1ex]

\emph{$x$-gadget:} By using the result of Theorem~\ref{Theorem: Stability cycle Form} for the green agent's subset $U^g = \{x_{t,1}, x_{t,2}, x_0\}$, observe that in any rejection-proof solution, the set of cycles and chains containing vertices from $U^g$ should cover at least 3 green vertices. This can only be realized through selecting either the cycle $c_{x,t} = \{x_{t,1}, x_{t,2}, x_0\}$ or the cycle $c_{x,f} = \{x_{f,1}, x_{f,2}, x_0\}$. Without loss of generality, suppose we select $c_{x,t}$.
Let now $S$ be a maximum rejection-proof solution, and suppose its sub-packing $S[G_x]$ induced by the $x$-gadget is not $S_{x,t}$. Again without loss of generality, suppose the cycle $\{x_{f,1}, \alpha_{x,1}, \gamma_{x,f,1}\}$ is not contained in $S[G_x]$. As $S$ is maximum, this means that the cycle $\{\gamma_{x,f,1}, \delta_{c(\neg x, 1)}\} \in S$ (if not, then all vertices of the cycle $\{x_{f,1}, \alpha_{x,1}, \gamma_{x,f,1}\}$ are uncovered in $S$). However, we could then switch from $\{\gamma_{x,f,1}, \delta_{c(\neg x, 1)}\}$ to $\{x_{f,1}, \alpha_{x,1}, \gamma_{x,f,1}\}$ without violating subset rejection constraints to obtain rejection-proof solution $S'$ with $|S'| = |S| + 1$, thus contradicting that $S$ is of maximum size.\newline

\emph{$y$-gadget:} Notice that any feasible rejection-proof solution $S$ has the following property: for each $y \in Y$, the sub-packing induced by $S[G_y]$ is maximal with respect to $G_y$. (If not, there exists a 3-cycle $\tilde{c}$ whose vertex set $U^b = V(\tilde{c})$ defines a violated subset rejection constraint). This can only be attained in four ways: the first two correspond to the gadget packings $S_{y,t}$ and $S_{y,f}$ consistent with $\true$ and $\false$. The other alternative maximal packings on $G_y$ are

\[S_3 = \Big\{\{\gamma_{y,f,1}, y_{f,1}, \alpha_{y,1}\}, \{\gamma_{y,t,2}, y_{t,2}, \alpha_{y,2}\}\Big\} \text{ and } S_4 = \Big\{\{\gamma_{y,t,1}, y_{t,1}, \alpha_{y,1}\}, \{\gamma_{y,f,2}, y_{f,2}, \alpha_{y,2}\}\Big\}.\]

Without loss of generality, suppose the $y$-gadget is packed according to $S_3$. Then the subset rejection constraint is violated for $U^b = \{y_{t,1}, y_{t,2}, y_0, \alpha_{y,2}, y_{f,2}, \gamma_{y,f,2}\}$, since $\beta(U^b) = 6 > 5$, which is the maximum number of vertices that could be covered in $S$ with cycles having nonempty vertex intersection with $U^b$ (see Figure~\ref{fig:solution_ygadget_zigzag} for an illustration). This shows that in a feasible rejection-proof solution, all $y$-gadgets must be packed consistent with either $\true$ or $\false$.

Since both cases cover all of the variable gadgets, the statement follows.
\end{proof}

Notice now that whenever we select a variable gadget packing consistent with a truth setting $\psi$, then the vertices $\gamma_{z,\psi,i}, i = 1,2$ are not covered by a cycle internal to the gadget. Nevertheless, we can use the corresponding cycles $(\gamma, \delta)$-cycles to cover these vertices. In that case, the $\delta$-vertex should not yet be covered by another $(\gamma, \delta)$-cycle. Observe that using a $(\gamma, \delta)$-cycle corresponding to a variable $z \in X \cup Y$ and clause $c \in C$ corresponds to satisfying the clause $c$ as the variable $z$ occurs in sense $\psi$ in clause $c$. \newline

\begin{figure}[ht!]
            \centering
            \begin{subfigure}[b]{0.49\linewidth}
                \centering
                \includegraphics[scale = 0.9]{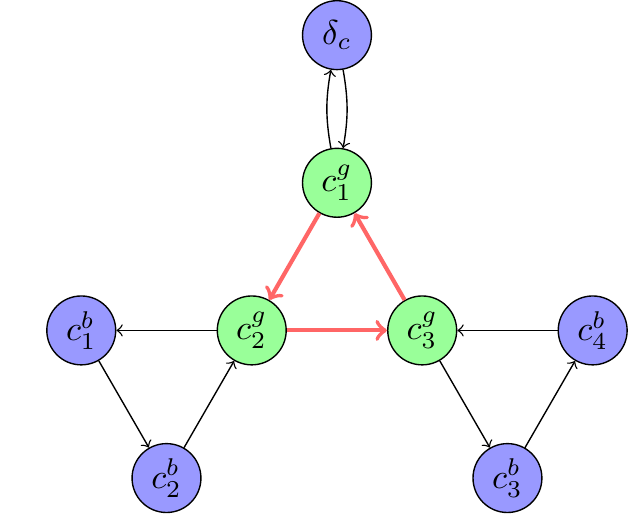}
                \caption{Satisfied clause packing $S_{c,t}$\label{fig:satisfied_clause}}
            \end{subfigure}
            \hfill
            \begin{subfigure}[b]{0.49\linewidth}
                \centering
                \includegraphics[scale = 0.9]{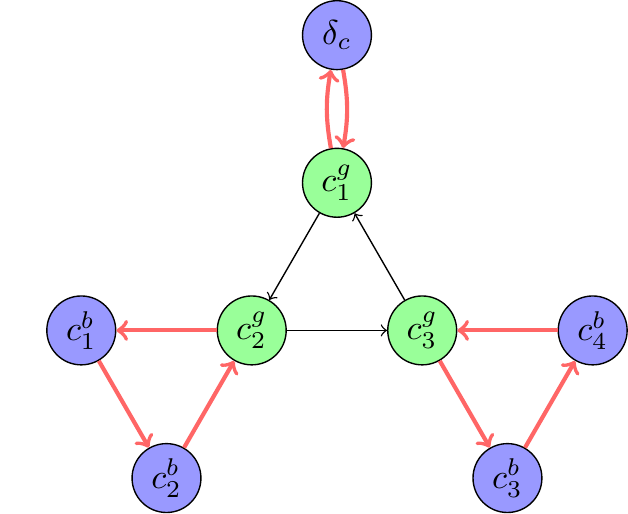}
                \caption{Unsatisfied clause packing $S_{c,f}$\label{fig:unsatisfied_clause}}
            \end{subfigure}
            \caption{Illustration of the satisfied clause packing  and the unsatisfied clause packing. As rejection-proofness requires all three green vertices are covered, these are the only possible packings in a feasible solution.\label{fig:solution_satgadget}}
        \end{figure}

For each clause $c \in C$, there are two possible packings for the clause gadget of $c$ based on whether or not the vertex $\delta_c$ is covered by a $(\gamma,\delta)$-cycle, in other words, whether or not clause $c$ is satisfied. If yes, we select the packing $S_{c,t} = \Big\{\{c_1^g, c_2^g, c_3^g\}\Big\}$, given by the unique green 3-cycle. Otherwise, the packing $S_{c,f} =\Big\{ \{\delta_c, c_1^g\}, \{c_2^g, c_1^b, c_2^b\}, \{c_3^g, c_3^b, c_4^b\}\Big\}$ is selected. We refer to $S_{c,t}$ and $S_{c,f}$ as the {\em satisfied clause packing} and the {\em unsatisfied clause packing} respectively. Both packings are depicted in Figure~\ref{fig:solution_satgadget}.
\begin{lemma}\label{lemma:clause_packing}
For each feasible solution to \emph{MaxRP-KEP} on $f(\varphi)$, each satisfiability gadget is packed with either $S_{c,t}$ or $S_{c,f}$.
\end{lemma}
\begin{proof}
We again use the result of Theorem~\ref{Theorem: Stability cycle Form} to show the statement. Consider the subset $U^g = W_c^g = \{c_1^g, c_2^g, c_3^g\}$ for some clause $c \in C$. Any packing in the clause gadget of $c$ should cover at least all vertices of $U^g$, and the only two packings with this property are given by $S_{c,t}$ and $S_{c,f}$.
\end{proof}

The following lemma illustrates the link between variable and clause gadgets. Whenever the variable gadgets correspond to a truth setting satisfying a clause, then there will be a $(\gamma,\delta)$ cycle selected, linking the clause gadget to a variable gadget. A satisfied clause gadget packing will result.
\begin{lemma}\label{lemma:satisfy}
Let the variable gadgets be packed consistent with a truth setting $\theta_X, \theta_Y$. In every feasible rejection-proof solution to $f(\varphi)$, for each clause $c \in C$ satisfied by this truth setting, $\delta_c$ is covered by a $(\gamma,\delta)$-cycle.
\end{lemma}
\begin{proof}
Let $c \in C$ be a clause satisfied with respect to truth settings $\theta_X, \theta_Y$. Without loss of generality, let $x \in X$ such that $c$ contains the first occurrence of the literal $x$ and $\theta_X(x) = \true$. Then, consider the blue agent's subset $U^b = \{\gamma_{x,t,1}, \delta_c\}$. If $\delta_c$ is not covered by a $(\gamma, \delta)$-cycle, then $U^b$ is completely uncovered while it induces a 2-cycle internal to blue. Hence, if $\delta_c$ is not covered by a $(\gamma, \delta)$-cycle, then the overall packing would not be rejection-proof, leading to a contradiction.
\end{proof}

Given a feasible solution to MaxRP-KEP on the instance $f(\varphi)$ in which every variable gadget is packed consistent with a truth setting, we define $\theta_X \colon X \to \{\true, \false\}$ as the truth setting of $X$ such that the variable gadget of $x \in X$ is packed by $S_{x,\theta_X(x)}$. Furthermore, let $k_{max}(\theta_X)$ be the maximum number of clauses in $C$ that can be satisfied given the truth setting $\theta_X$ for $X$, over all truth settings for $Y$.

\begin{lemma}\label{lemma:satisfying_TA}
Let $S$ be a feasible solution to MaxRP-KEP on $f(\varphi)$ such that $\theta_X$ is the associated truth setting for $X$. If $k_{max}(\theta_X) = |C|$, then the packing of each $y$-gadget in $S$ is $S_{y,\theta_Y(y)}$ for some truth setting $\theta_Y$ for $Y$ that satisfying all clauses.
\end{lemma}
\begin{proof}
Suppose by contradiction that the packing of $y$-gadgets in $S$ is not equal to $S_{y,\theta_Y(y)}$ for some satisfying truth setting $\theta_Y$ for $Y$ given $\theta_X$, i.e. using $\theta_X$ and $\theta_Y$ together results in $k < k_{max}(\theta_X) = |C|$ satisfied clauses. We will show the blue agent has incentive to deviate.
We consider the subset $U^b \subset V^b$ of pairs of the blue agent defined by
\[ U^b \coloneqq (\bigcup_{y \in Y} V_y^b) \cup \{\delta_c \mid c \in C\} \cup \{\gamma_{x,\psi,i} \mid x \in X, \psi \in \{t,f\} \colon \theta_X(x) = \psi\} \]

Currently, it holds for solution $S$ that
\[ \sum_{e \in S: V(e) \cap U^b \ne \emptyset} w^b_e = \sum_{e \in S: V(e) \subseteq U^b} w^b_e = 9|Y|+|C|+k.\]
Furthermore, since $k_{max}(\theta_X) = |C|$, we have that $\beta(U^b) = 9|Y| + 2|C| > 9|Y| + |C| + k$, as each variable gadget allows for an internal packing covering 9 vertices, and satisfiability of all clauses implies that we can pick $|C|$ disjoint $(\gamma, \delta)$-cycles internal to $U^b$ by construction of the gadgets and the structure of $U^b$. Hence, the subset rejection constraint for agent $b$ and subset $U^b$ is violated at solution $S$, which contradicts that $S$ is a rejection-proof solution.
\end{proof}

We use Lemma~\ref{lemma:consistent_var_pack} to~\ref{lemma:satisfying_TA} to prove our main complexity result, Theorem~\ref{th:complexity_proof}.
\begin{theorem*}\label{thm:poly_reduct}
DecRP-KEP is $\Sigma_2^p$-complete, for fixed maximum cycle length $K \ge 3$ and arbitrary maximum chain length $L \in \N$.
\end{theorem*}
\begin{proof}
We start by showing membership of DecRP-KEP in $\Sigma_2^p$. Following the definition of $\Sigma_2^p$ in~\cite{Arora2009}, we consider DecRP-KEP as the language
\begin{align*}
    \text{DecRP-KEP} = \{\langle G, \P, K, L, t \rangle : &\exists \text{ rejection-proof KEP solution consisting of cycles of length $\le K$}\\
    & \text{and chains of length $\le L$ covering at least $t$ vertices.}\}
\end{align*}

Remark that this language is in $\Sigma_2^p$ since a tuple $\langle G, \P, K, L, t \rangle \in$ DecRP-KEP if and only if
\begin{align*}
    &\exists X \in \X:\ X \text{ covers at least $t$ vertices.}\\
    &\forall X' \in \X:\  X' \text{ is either (i) not feasible to RKEP$(X,i)$ or (ii) has value $w^i(X') \le w^i(X)$ for all $i \in \P$.}
\end{align*}
Remark that there exist compact encodings of $X, X'$, i.e. polynomial in number of vertices and arcs of the graph $G$, such as arc-based formulations of kidney exchange programs. Furthermore, it can be checked in polynomial time whether such encodings indeed imply feasible cycle-chain packings. This also holds for checking if $X'$ is feasible to RKEP$(X, i)$ and checking its corresponding objective value $w^i(X')$. Therefore, given a rejection-proof solutions $X$ and some alternative vector $X'$, we can check in polynomial time that $X'$ does not prove violation of rejection-proofness of $X$, hence the language DecRP-KEP is in $\Sigma_2^p$.

In order to prove $\Sigma_2^p$-hardness, we use the results of Lemma~\ref{lemma:consistent_var_pack} to~\ref{lemma:satisfying_TA}. The reduction is based on proving the following statement: an instance $\varphi$ to Adversarial (2,2)-SAT returns YES if and only if the transformed instance $f(\varphi)$ to DecRP-KEP returns YES. Recall that we use the parameter settings $\P = \{b,g\}, K = 3$, $L \in \mathbb{N}$, and $t = 9|X|+9|Y|+5|C|+3$.

$\Rightarrow)$ Let us be given a YES-instance $\varphi = \varphi(X,Y,C)$ to Adversarial (2,2)-SAT, i.e. there exists a truth setting $\theta_X \colon X \to \{\texttt{True, False}\}$ such that for any truth setting $\theta_Y \colon Y \to \{\texttt{True, False}\}$, there exists some clause $c \in C$ that is left unsatisfied. Let us now fix such a truth setting $\theta_X$ for $X$. Let $k_{max}(\theta_X) \le |C|-1$ be the maximum number of clauses of $C$ that can be satisfied given this $\theta_X$, and let $\theta_Y$ be a truth setting attaining $k_{max}(\theta_X)$ satisfied clauses. We construct a solution $S$ to the corresponding DecRP-KEP instance $f(\varphi)$ as follows:
\begin{itemize}
    \item for each variable $x \in X$, we add to $S$ the packing $S_{x, \theta_X(x)}$ of the $x$-gadget consistent with $\theta_X(x)$.
    \item similarly, for each variable $y \in Y$, we add to $S$ the packing $S_{y, \theta_Y(y)}$ of the $y$-gadget consistent with $\theta_Y(y)$
    \item for each clause $c \in C$, we add to $S$ the satisfied clause packing $S_{c,t}$ if $c$ is satisfied by the truth settings $\theta_X$ and $\theta_Y$, and the unsatisfied clause packing $S_{c,f}$ otherwise.
    \item for each $(\gamma, \delta)$-cycle, add it to $S$ if both its vertices are not covered yet by $S$.
\end{itemize}
Within each $X$- and $Y$-gadget, the solution $S$ covers nine vertices, which gives us $9|X|+9|Y|$ vertices already.
Furthermore, since the corresponding truth settings $\theta_X$ and $\theta_Y$ for $X$ and $Y$ respectively ensure that $k_{max}(\theta_X)$ clauses were satisfied, this gives us another $k_{max}(\theta_X)$ cycles of length 2, so an additional $2k_{max}(\theta_X)$ vertices are covered.
Finally, if the $c$-vertex was not in a 2-cycle already, we cover eight vertices internally in the satisfiability gadget, opposed to 3 otherwise.
In total, this gives us
\[9|X|+9|Y| + 2k_{max}(\theta_X) + 8(|C|-k_{max}(\theta_X))+3k_{max}(\theta_X) = 9|X|+9|Y| + 8|C| - 3k_{max}(\theta_X) \]
vertices covered by $S$. This is indeed at least $9|X|+9|Y|+5|C|+3$, since $k_{max}(\theta_X) \le |C|-1$.
It remains to show that this solution is indeed rejection-proof. The transformed instance $f(\varphi)$ of DecRP-KEP contains a green and blue agent. Observe that $S$ covers all green vertices, so we can restrict ourselves to proving that the blue agent has no incentive to reject $S$. First, notice that the blue agent cannot use an internal exchange to cover the vertices $c_i^b, i \in \{1,2,3,4\}$ in the clause gadgets. Therefore, such vertices can only be covered if the proposed solution includes its associated shared cycle. Furthermore, proposed shared cycles within each gadget of variable $z \in X \cup Y$ of the form
$\{z_{\psi,i}, \alpha_{z,i}, \gamma_{z, \psi, i}\}$, where $\psi \in \{t,f\}$ and $i \in \{1,2\}$, will also not be rejected by the blue agent. Notice that the only reason to reject such a cycle is to be able to pick the $(\gamma, \delta)$-cycle incident with $\gamma_{z, \psi, i}$. However, there is no strict benefit to the blue agent to make this switch. Furthermore, this blocks the $\delta$-vertex from forming a 2-cycle with another blue $\gamma$-vertex from another variable gadget. Therefore, accepting the cycle always leads to weakly preferred outcomes to the blue agent.
The remaining set of the blue agent's vertices is $U^b$ as defined in the proof of Lemma~\ref{lemma:satisfying_TA}. Notice that in $S$, all $\delta_c$-vertices for $c \in C$ are covered. Furthermore, we have maximum size packings within $x$- and $y$-gadgets that allow for the largest number, i.e. $k_{max}(\theta_X)$, of $(\gamma, \delta)$-cycles, with respect to the set of solutions that packs all $x$- and $y$-gadgets consistent with a truth setting. This means that a total number of $9|Y|+|C|+k_{max}(\theta_X)$ vertices in $U^b$ are covered by a cycle in $S$. Hence, a certificate solution $S'$ showing that the blue agent has incentive to reject $S$ needs to contain more $(\gamma, \delta)$-cycles than $S$. Recall that $S$ corresponds to a truth setting for $Y$ that maximizes the number of satisfied clauses. Thus, there must exist a variable $y \in Y$ such that $S'$ contains a $(\gamma, \delta)$-cycle associated with occurrences of both the unnegated and negated literal of $y$. Whenever this happens, the packing of the $y$-gadget is not consistent anymore and a 3-cycle within the $y$-gadget needs to be given up, since this would mean a $\gamma_y$-vertex is covered with two distinct cycles. Therefore, a solution containing $k - k_{max}(\theta_X)$ extra $(\gamma, \delta)$-cycles comes at the cost of giving up equally many 3-cycles within the $y$-gadgets, hence covering fewer vertices of $U^b$ in total compared to $S$. This shows that $S$ is rejection-proof and satisfies the target $t$, i.e. $S$ is a certificate solution proving that $f(\varphi)$ is a YES-instance to DecRP-KEP.



$\Leftarrow)$ Let us consider a feasible solution $S$ to DecRP-KEP for the transformed instance that covers at least $9|X|+9|Y|+5|C|+3$ vertices, meaning that $S$ is in fact of maximum size. Remark that there must exist a clause $c \in C$ for which the $c$-gadget is packed with $S_{c,f}$, since otherwise a feasible DecRP-KEP solution can have at most $9|X|+9|Y|+2|C|+3|C| = 9|X|+9|Y|+5|C|$ covered vertices otherwise. Let the truth setting $\theta_X \colon X \to \{\texttt{True, False}\}$ such that the variable gadgets for each $x \in X$ are packed in $S$ by $S_{x, \theta_X(x)}$. This follows from Lemma~\ref{lemma:consistent_var_pack}, which assures us that each variable gadget packing is consistent with a truth value. Since at least one unsatisfied clause packing is selected and $S$ is rejection-proof, Lemma~\ref{lemma:satisfying_TA} implies that there cannot exist a truth setting to $Y$ that satisfies all clauses given $\theta_X$. Therefore, $\theta_X$ is a valid certificate for $\varphi = \varphi(X,Y,C)$ being a YES-instance to Adversarial (2,2)-SAT.
\end{proof}

\section{Computation times}
\subsection{Difficulty of instance sets} \label{sec:difinstancesets}
While the basic characteristics of the instances such as number of pairs and NDDs were kept constant, we note massive differences in computation time required for the different instance sets. These differences were most pronounced for the different distributions used for instance generation (Saidman vs. SplitPRA-BandXMatch-PRA0), with the Saidman instances considerably more difficult. Agent preferences also had a large impact, as instances with unit agent valuations for both the overall objective and agent preferences were significantly easier to solve. Finally, more minor differences exist depending on the distribution of pairs and NDDs over the agents.

\begin{itemize}
    \item \textbf{Saidman vs. Delorme.} It is clear that there exist huge differences based on the instance generation procedure used. All instance sets generated using the Delorme distributions are solved completely, with average computation times only exceeding 1 second for one set of instances. Instances generated using the Saidman distributions on the other hand are much more difficult, with not all instances solved within the time limit. For the solved instances, the average time to optimality is also much longer, up 12 minutes for one instance set.

    A likely factor for this difference is the far greater arc density in Saidman instances (see \cite{delorme2021new}) For one, this increases the computational cost of solving the integer programs. Second, this gives the agents more freedom and they are thus more likely to reject proposed solutions, as reflected in the number of iterations required to find the optimal rejection-proof solution.

    \item \textbf{Unit vs. Score} Instances in which the agents have non-unit values for the transplants are much more difficult. This very apparent in the Saidman instances, where all instances with unit valuations are solved within the time limit, with the most difficult instances set taking less than 6 seconds on average. For the non-unit values, instances in each set were unsolved, and average times for solved instances range from 5 to 12 minutes.

    The reason for the additional difficulty is reflected in the number of iterations and number of constraints. In case of unit values, the overall value and agent values of a given solution are closely aligned. For non-unit values however, solutions with many low value transplants are often rejected by the agents, adding more constraints to the master problem. This leads to more, and more costly, iterations as the additional constraints add difficulty to the master problem.
    \item \textbf{Distribution over agents} We focus on the difficult Saidman Score instances, and note instances with larger agents are more difficult to solve. The ``4equal" set has most instances solved, and while the average time per instance is second highest, this is due to the final solved instances taking just under 3 hours. Excluding this instance, the average stands at only 163 seconds. For all other instances, the largest agent has 50 pairs and NDDs, and we see that more instances are solved, with lower average time, for the instances were the remaining pairs are divided over more agents.

    Since larger agents have more opportunities to identify profitable rejections, the increased difficulty is as expected. This is also reflected in the number of iterations required and constraints added. In instance sets with larger agents, more constraints were needed to identify rejection-proof solutions.
\end{itemize}

The instance sets in this study cannot capture the variety of possibilities possible in real life, and the feasibility of using MaxRP-KEP in practice will depend on the specific properties of the application. This said, we believe the above results are positive. MaxRP-KEP is solvable on most medium-size instances even for the dense Saidman instances, and trivial for the more realistic Delorme instances. The increased difficulty as agent size increases is a concern, especially in cooperations between large countries. Still, size 50 agents are manageable, a size which is only exceeded by 3 European Kidney Exchanges (UK, Spain and The Netherlands \citep{biro2019}).

\subsection{Tiebreaker}
We now discuss the tiebreaker, and its effect on computation time for the challenging SaidmanScore instances. Its effect depends on instance characteristics. Figure \ref{Fig:tiebreak} shows the number of solved instances at each point in time and gives the clearest comparison. For the instances with one large and 2 medium or 5 small agents, the tiebreaker generally speeds up computation for easier instances, but the gap disappears for harder instances. For both instance sets, not using the tiebreaker in fact allows for one more instance to be solved within the time limit. For the 2equal and 4equal instance sets, little difference can be seen. We do note that using the tiebreaker strongly reduces the number of iterations required to identify optimal solutions, as can be seen in Table \ref{Table: tiebreak}. As before, this is offset by the increased difficulty of solving the master problem, now due to the harder objective function, leading to similar overall results.

\begin{table}[htbp]
    \centering\footnotesize
    \begin{tabular}{llrrrr}
        \toprule
        Instance Set & tiebreak &  Total Time (s) &  Master Time (s) &  Iterations &  Instances Solved \\
        \midrule
        1l2m & Used &          308.04 &           298.82 &       22.83 &                24 \\
                                 & Not Used &          942.85 &           911.45 &       44.48 &                25 \\
        1l5s & Used &          303.07 &           290.88 &       28.67 &                27 \\
                                 & Not Used &          726.00 &           672.74 &       43.71 &                28 \\
        2equal & Used &          732.42 &           720.24 &       28.04 &                24 \\
                                 & Not Used &         1176.52 &          1156.51 &       39.17 &                24 \\
        4equal & Used &          518.13 &           516.44 &       13.89 &                28 \\
                                 & Not Used &          378.73 &           376.41 &       18.71 &                28 \\
        \bottomrule
    \end{tabular}
    \caption{Comparison of SaidmanScore instances based on whether or not tiebreaks are used. Results were obtained using no initial constraints and maximum violated constraints added. Instances solved is the number of instances solved to optimality within 3 hours (10800s), out of 30. Reported computation and master times, and numbers of iterations are averages for instances solved to optimality within this time.}
    \label{Table: tiebreak}
\end{table}
\end{appendices}

\end{document}